\documentclass[a4paper,12pt]{article}
\usepackage{graphicx}
\usepackage{amsmath}
\usepackage{multirow}
\usepackage{float} 
\usepackage{amssymb}
\usepackage{cite}

\begin{document}

\begin{center}
{\large\bf Prospects of Neutral Higgs boson decays in the NMSSM } \\
\vspace*{1.0truecm}
{\large M. M. Almarashi} \\
\vspace*{0.5truecm}
{\it Departement of Physics, Taibah University, Almadinah, KSA}
\end{center}

\begin{abstract}
In this paper, we present the branching ratios of the two lightest CP-even Higgses, $h_1$ and $h_2$, and of the lightest CP-odd Higgs, $a_1$,
in the context of the Non-Minimal Supersymmetric Standard Model (NMSSM). We find that for some regions of the parameter space, the branching
fractions of $a_1$ into the lightest charginos ($a_1\to  \chi^+_1\chi^-_1)$ and neutralinos ($a_1\to \chi^0_1\chi^0_1)$ are larger than $50\%$,
and can reach a $100 \%$ level in the latter case. Furthermore, the branching ratio of the $h_2\to a_1a_1$ can be equal to unity
for all masses of $h_2$ when it is a pure singlet. The above unusual channels are hallmarks of 
the existence of the NMSSM.

\end{abstract}

\section{Introduction}

The discovery of a SM-like Higgs particle with a mass around 125 GeV by ATLAS \cite{ATLAS1-Higgs,ATLAS2-Higgs} 
and CMS \cite{CMS1-Higgs,CMS2-Higgs} is a great success of
the Standard Model (SM) \cite{SM1,SM2,SM3}
and Supersymmetry (SUSY) \cite{SUSY1,SUSY2,SUSY3,SUSY4,SUSY5,SUSY6,SUSY7,SUSY8}. However, from the theoretical point of view the SM
can not be the ultimate theory of particle physics as
it leaves some problems unsolved such as the hierarchy problem: the huge gap between the unification scale and electroweak scale,
no existence of good dark matter candidate, no explanation of Gravity and 
so on. Not only the SM has problems but also the Minimal Supersymmetric Standard
Model (MSSM) \cite{MSSMreview} has other problems such as  suffering from the $\mu$-problem \cite{Kim:1983dt} and the naturalness issue
\cite{Weinberg:1978ym,LlewellynSmith:1981yi}, so extensions beyond the MSSM have been introduced.
One of those good extensions that accommodates the measured value of the SM-like Higgs boson mass
and solves the above MSSM problems is the Next-to-Minimal Supersymmetric Standard Model (NMSSM)
\cite{NMSSMreviewed1,NMSSMreviewed2}. Many studies have been done
on the NMSSM phenomenology with Higgs mass around 125 GeV \cite{NMSSMrecent1,NMSSMrecent2,NMSSMrecent3,NMSSMrecent4,
NMSSMrecent5,NMSSMrecent6,NMSSMrecent7,NMSSMrecent8,NMSSMrecent9,NMSSMrecent10,NMSSMrecent11,NMSSMrecent12,NMSSMrecent13,NMSSMrecent14}.
Although there is no direct evidence for the existence of the supersymmetric particles, the NMSSM is still phenomenologically interesting, 
and many theoretical and experimental scientists
have contributed to the model through their recent works, see, e.g., Refs. \cite{recentstudy1,recentstudy2,recentstudy3,recentstudy4,recentstudy5,recentstudy6}.

The NMSSM is phenomenologically richer than the MSSM in the Higgs and neutralino sectors. This is
due to introducing an extra complex singlet 
superfield, which only couples to the two MSSM-type Higgs doublets. Therefore, the spectrum of the NMSSM compared to the MSSM contains
one more CP-even Higgs, one more CP-odd Higgs and one more neutralino. As for the neutralino sector of the NMSSM,
there is a possibility that the lightest neutralino, which has a singlino component, is a dark matter candidate.

In the NMSSM, the soft Supersymmetry breaking potential of the Higgs sector is given by the
following contribution
\begin{equation*}
V_{\rm NMSSM}=m_{H_u}^2|H_u|^2+m_{H_d}^2|H_d|^2+m_{S}^2|S|^2
             +\left(\lambda A_\lambda S H_u H_d + \frac{1}{3}\kappa A_\kappa S^3 + {\rm h.c.}\right),
\end{equation*}
where $H_u$ and $H_d$ are the Higgs doublet fields, $S$ is the Higgs singlet field, $\lambda$ and $\kappa$  are dimensionless
couplings while $A_\lambda$ and $A_\kappa$ are dimensionful parameters of order $M_{\rm{SUSY}}$,
the SUSY mass scale. After Electroweak Symmetry Breaking (EWSB), the neutral components of the Higgs doublet fields acquire
Vacuum Expectation Values (VEVs) giving rise to 
seven physical Higgs bosons: three CP-even Higgses $h_{1, 2, 3}$ ($m_{h_1} < m_{h_2} < m_{h_3}$), 
two CP-odd Higgses $a_{1, 2}$ ($m_{a_1} < m_{a_2} $) and a pair of charged Higgses $h^{\pm}$. 

In this paper, we will do a thorough study about the decay modes of the $h_1$, $h_2$ and  $a_1$, aiming to check the most important decay modes
of the neutral Higgs bosons, and study them in detail. Also, we aim to test if there exist some regions of
the NMSSM parameter space that impossible to exist in the MSSM. We will take into account only the lightest neutralinos $\chi^0_1\chi^0_1$
 and chargions $\chi^+_1\chi^-_1$ decay modes.

This work is organized as follows. In Sect.~\ref{sect:scan} we briefly describe our parameter space scans
. In Sect.~\ref{sect:Higgs-decays} we study the Higgs decay modes of the neutral Higgs bosons of the NMSSM and present
our results.
Finally, we summarize and conclude in Sect.~\ref{sect:summa}. 

\section{NMSSM Parameter Space Scan}
\label{sect:scan}
We perform a scan over large regions of the NMSSM parameter scan using the NMSSMTools package ~\cite{NMHDECAY1,NMHDECAY2,NMSSMTools}\footnote
{We have used NMSSMTools$\_$4.5.1.}. The package
computes the masses, couplings and decay widths of all the NMSSM Higgs
bosons. It also computes the masses of sparticles.
All the above calculations are computed in terms of the NMSSM parameters at the SUSY breaking scale.
NMSSMTools takes into account theoretical and
experimental constraints from negative Higgs searches at LEP and Tevatron 
including the unconventional channels relevant for the NMSSM. Furthermore, it takes into account some constraints coming from B-physics, constraints
on a SM-like Higgs mass and its signal rates.

Because of the large number of the NMSSM input parameters, it is practically unfeasible to make a continuous scan over all the parameter space.
The ranges of these parameters can be reduced significantly by assuming some conditions of
unification. Since the mechanism of the SUSY breaking is still unknown,
we have done a general scan through the following parameter space:
\begin{center}
$\lambda$ and $\kappa$: 0.01 --
0.7,\phantom{a} $\tan\beta$: 1.6 -- 60,\phantom{a} $\mu_{\rm eff}$: 100 -- 1000 GeV,\phantom{a} 
$A_{\lambda}$ and $A_{\kappa}$: -2000 -- 2000 GeV,\phantom{a} $M_1$: 100 -- 500 GeV ,\phantom{a} $M_2$: 100 -- 1000 GeV ,\phantom{a}
$M_3$: 100 -- 2000 GeV\footnote{Here, we treat the gaugino mass parameters $M_1$, $M_2$ and $M_3$ as free parameters and do not impose
any constraints on them as the masses of neutralinos and charginos depend on them.}. \\
\end{center}
The remaining scalar masses and trilinear parameters which have been fixed in the scan include:\\
$\bullet\phantom{a}m_{Q_3} = m_{U_3} = m_{D_3} = m_{L_3} = m_{E_3} = 1000$ GeV, \\
$\bullet\phantom{a}A_{U_3} = A_{D_3} = A_{E_3} = 2000$ GeV. 

We have performed a random scan over one million points in the NMSSM parameter space as mentioned above. The output of the scan is 
the surviving data points that have passed 
the experimental and theoretical constraints 
containing Branching Ratios (BRs), couplings of
the NMSSM Higgses, and the masses of Higgses and sparticles. The points that violate any type of the constraints
are automatically eliminated by the package.

\section{Higgs Decay Modes}
\label{sect:Higgs-decays}

According to the Ref. \cite{NMSSM-Points}, the possible decay 
channels for neutral NMSSM CP-even Higgs boson $h$, where $h=h_{1, 2, 3}$, and neutral CP-odd Higgs boson $a$,
where $a=a_{1, 2}$, are:
\begin{eqnarray*}
h,a\rightarrow gg,\phantom{aaa} h,a\rightarrow \mu^+\mu^-,
&&h,a\rightarrow\tau^+\tau^-,\phantom{aaa}h,a\rightarrow
b\bar b,\phantom{aaa}h,a\rightarrow t\bar t, \\ h,a\rightarrow
s\bar s,\phantom{aaa}h,a\rightarrow
c\bar c,&&h\rightarrow W^+W^-,\phantom{aaa}h\rightarrow ZZ, \\
h,a\rightarrow\gamma\gamma,\phantom{aaa}h,a\rightarrow
Z\gamma,&&h,a\rightarrow {\rm Higgses},\phantom{aaa}h,a\rightarrow
{\rm sparticles},
\end{eqnarray*}
where  
the `${\rm Higgses}$' denotes any possible final state Higgs particles
involving either two neutral Higgses, two charged Higgses, or one Higgs particle and one gauge vector boson, 
those combinations in the final state are allowed when they are compatible with the CP quantum number of 
the decaying Higgs boson. The above possible decay channels of the $h_{1, 2, 3}$ and $a_{1, 2}$ are only
exist above the respective kinematic thresholds.

As a first step, we have plotted the mass of $h_1$ as a function of its decay rates
as is shown in Fig. 1 \footnote{We do not study the channels $h,a\rightarrow s\bar s$, $h,a\rightarrow 
c\bar c$ and $h,a\rightarrow gg$ due to the smallness of their branching fractions in general.}.
It is clear from the figure that $h_1$ is always a SM-like Higgs boson with mass between $122$ and $128$ GeV
for all points
passed the constraints. Generally, such points correspond to small values of $\lambda$ (below 0.3) and $\mu_{\rm eff}$ (below 500), negative values of
$A_{\lambda}$ and  $A_{\kappa}$, uniform values of $\kappa$,  and low and middle values of $\tan\beta$. Nevertheless, there is regions 
of the parameter space where the very high values of $\tan\beta$ (above 30) are  
compatible with the current theoretical and
experimental constraints. For these large values of $\tan\beta$, the Higgs production in association with a $b$-quark pair is dominant and become
the best production mode for producing the NMSSM Higgs bosons at the LHC
\cite{Almarashi:2010jm,Almarashi:2011bf,Almarashi:2011hj,Almarashi:2011te,Almarashi:2011qq,NMSSMreviewed3}. 
The ratio of the
decays $h_1\to b\bar b$, $h_1\to \tau^+\tau^- $, $h_1\to \mu^+\mu^- $,
$h_1\to \gamma\gamma$, $h_1\to Z\gamma$ and $h_1\rightarrow\chi^0_1\chi^0_1$ are approximately $50 \%$ -- $75 \%$, $5 \%$ -- $8 \%$, 
$0.02 \%$-- $0.03 \%$, $0.1 \%$ -- $0.3 \%$, $0.1 \%$ -- $0.2 \%$ and $0.001 \%$ -- $30 \%$ respectively
\footnote{The decay channel $h_1\to a_1a_1$ will
be included in the paper later when we study the Higgs-to-Higgs decays.}.
From the various successful points showing in the figure, it is clear
that the NMSSM is a flexible model to explain the couplings of the SM-like Higgs, discovered at the LHC, to other particles. This depends on
the doublet and singlet components of the CP-even Higgs boson.  The bottom-right panel of
the figure shows that there are only few points of the parameter space where the $h_1$ decay to the lightest
neutralinos is kinematically allowed \footnote
{Since the branching ratios of Higgs into neutralinos and into charginos depend on their masses,
we will study the Higgs decays to the lightest neutralinos $\chi^0_1 \chi^0_1$ and to the lightest charginos $\chi^+_1 \chi^-_1$  
as they are the highest decay rates among the other neutralinos and charginos.}.
 
\begin{figure}
 \centering\begin{tabular}{ccc}
 \includegraphics[scale=0.50]{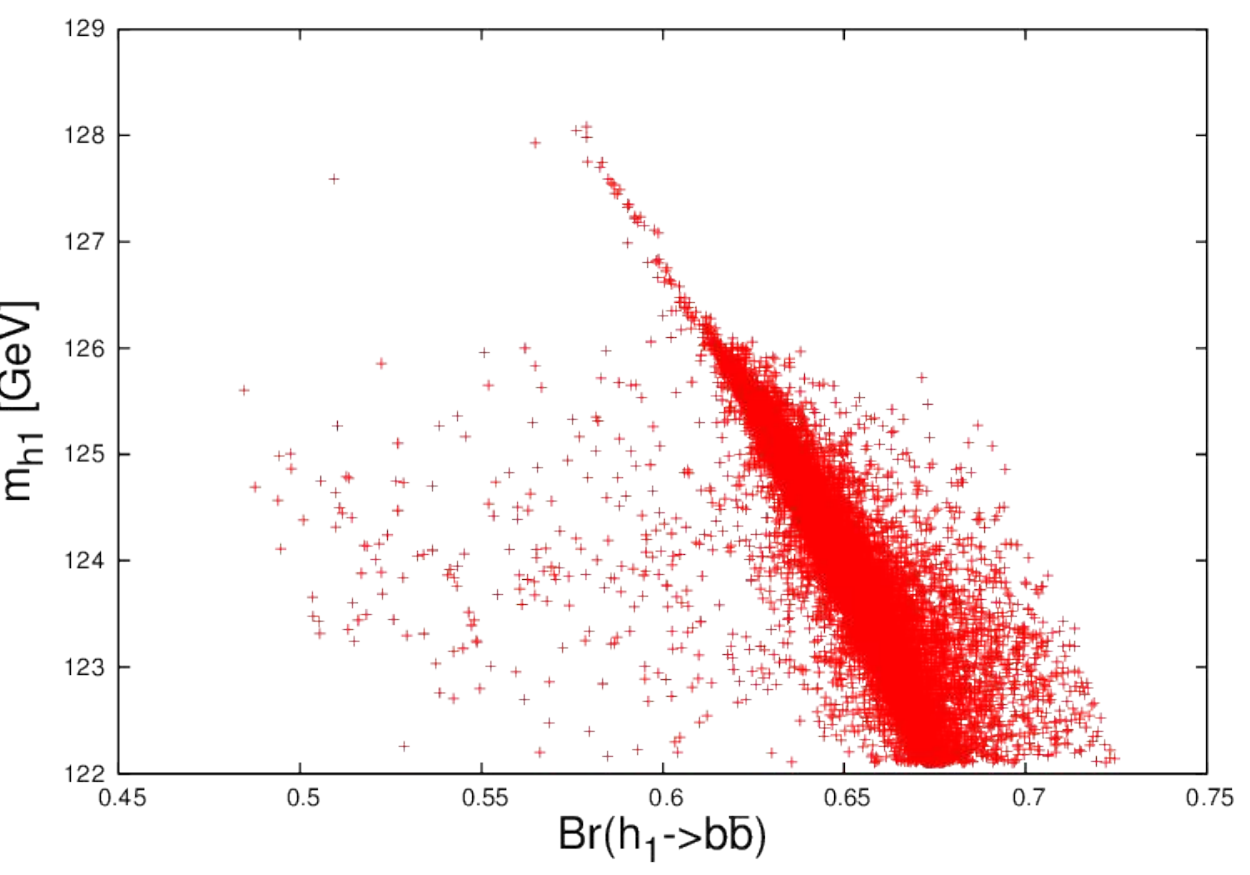}
 &\includegraphics[scale=0.50]{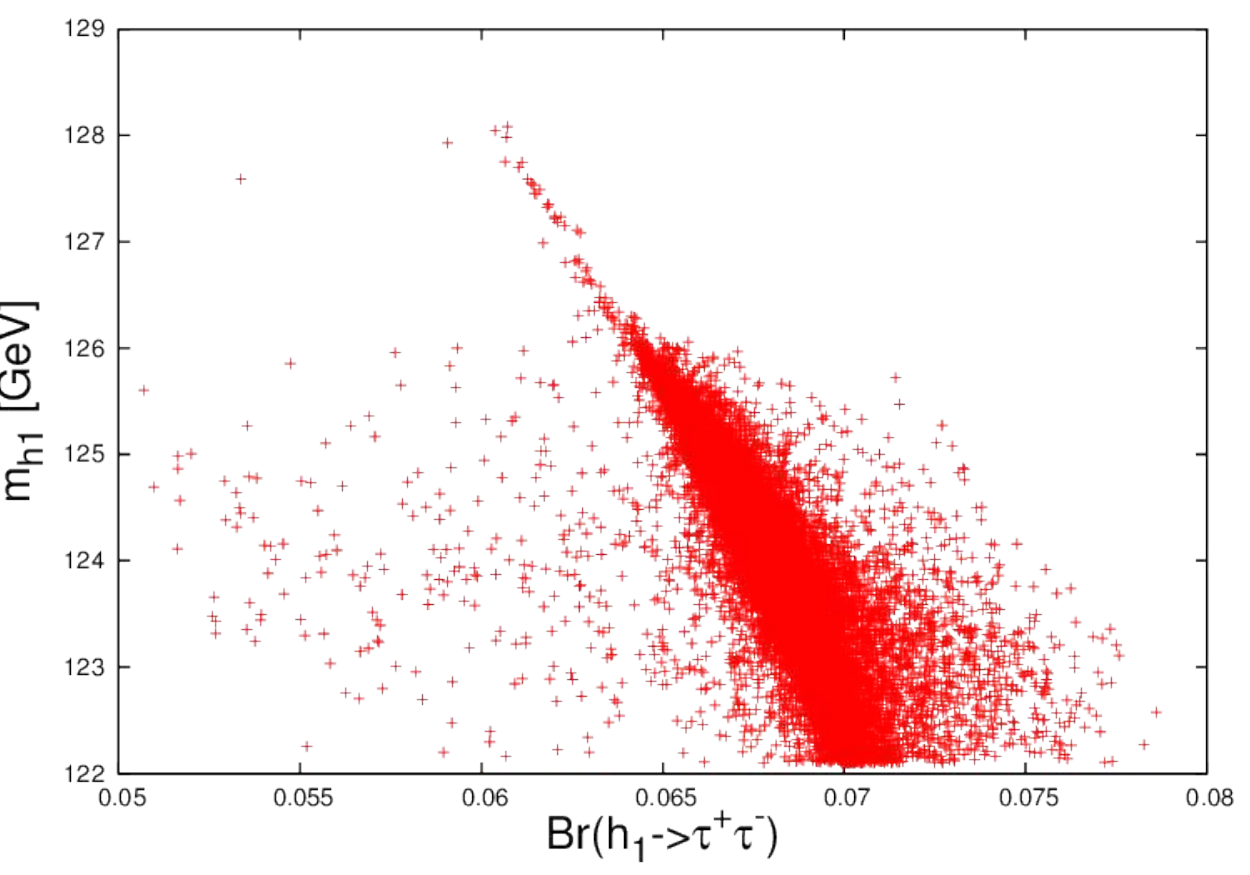}\\
 \includegraphics[scale=0.50]{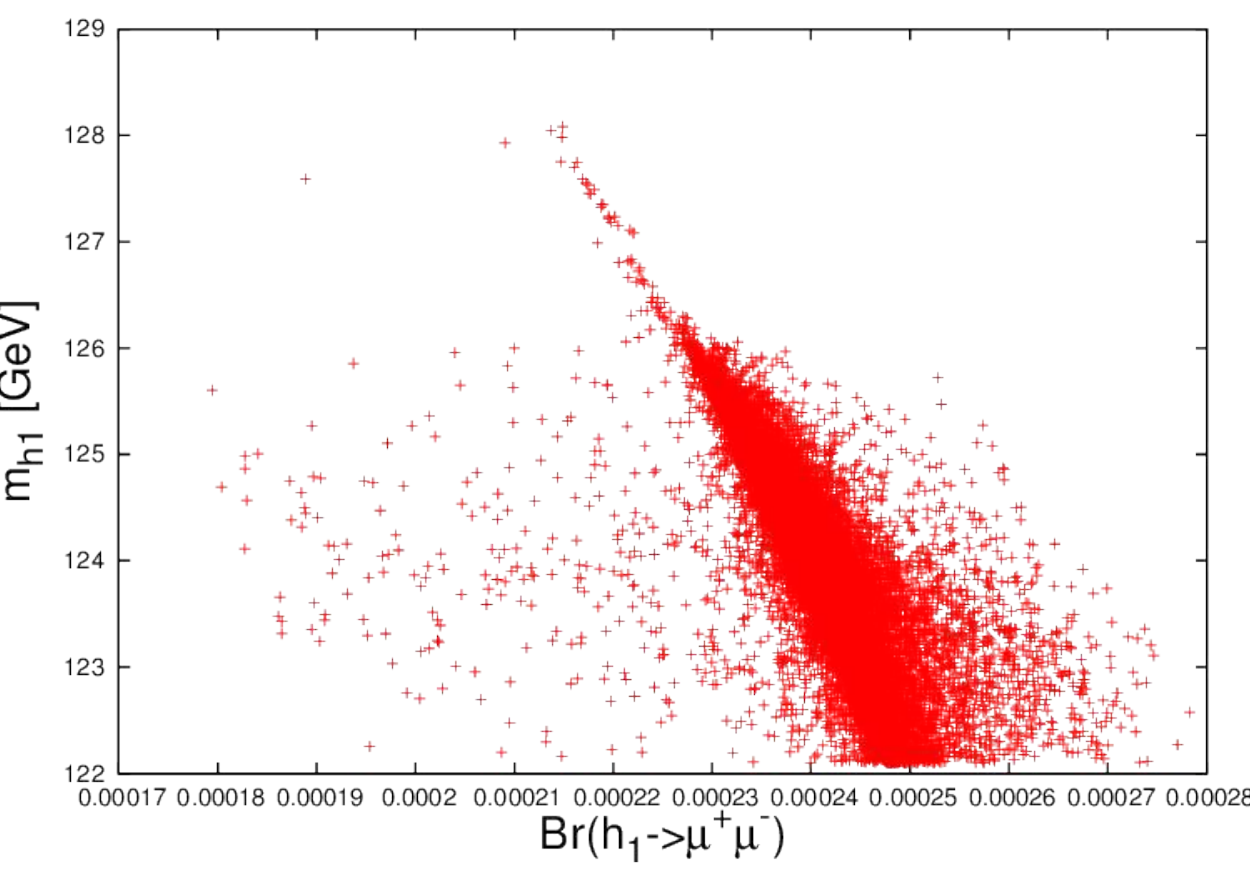}
 &\includegraphics[scale=0.50]{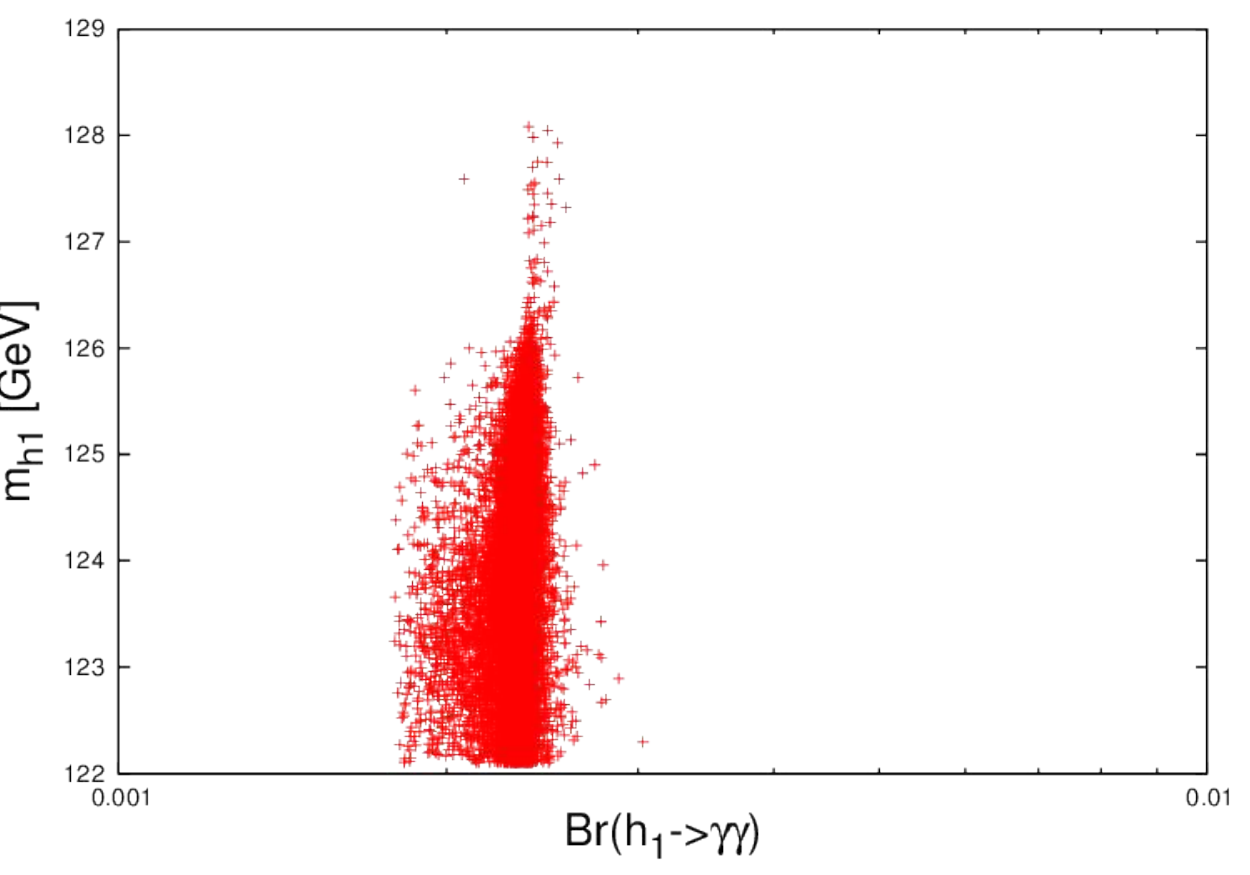}\\
 \includegraphics[scale=0.50]{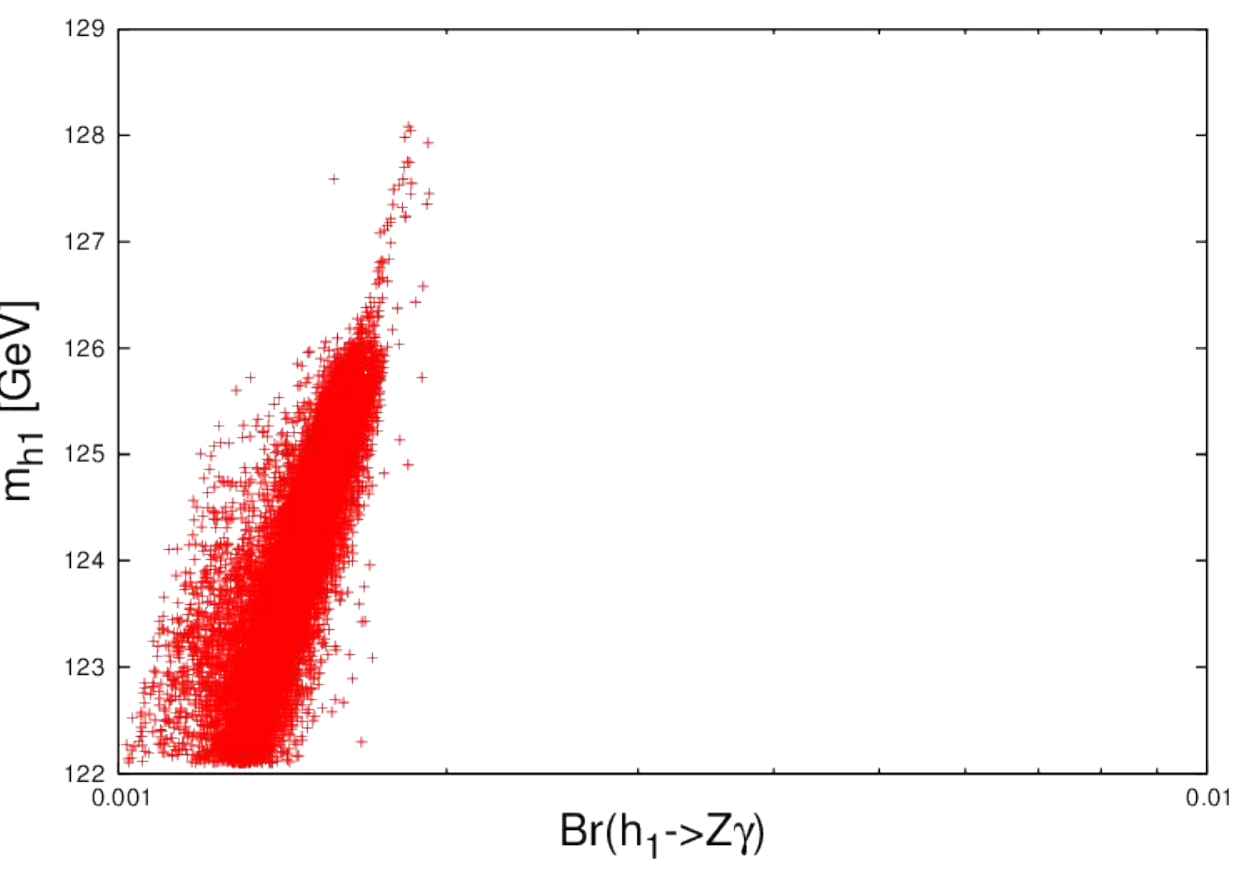}
 & \includegraphics[scale=0.50]{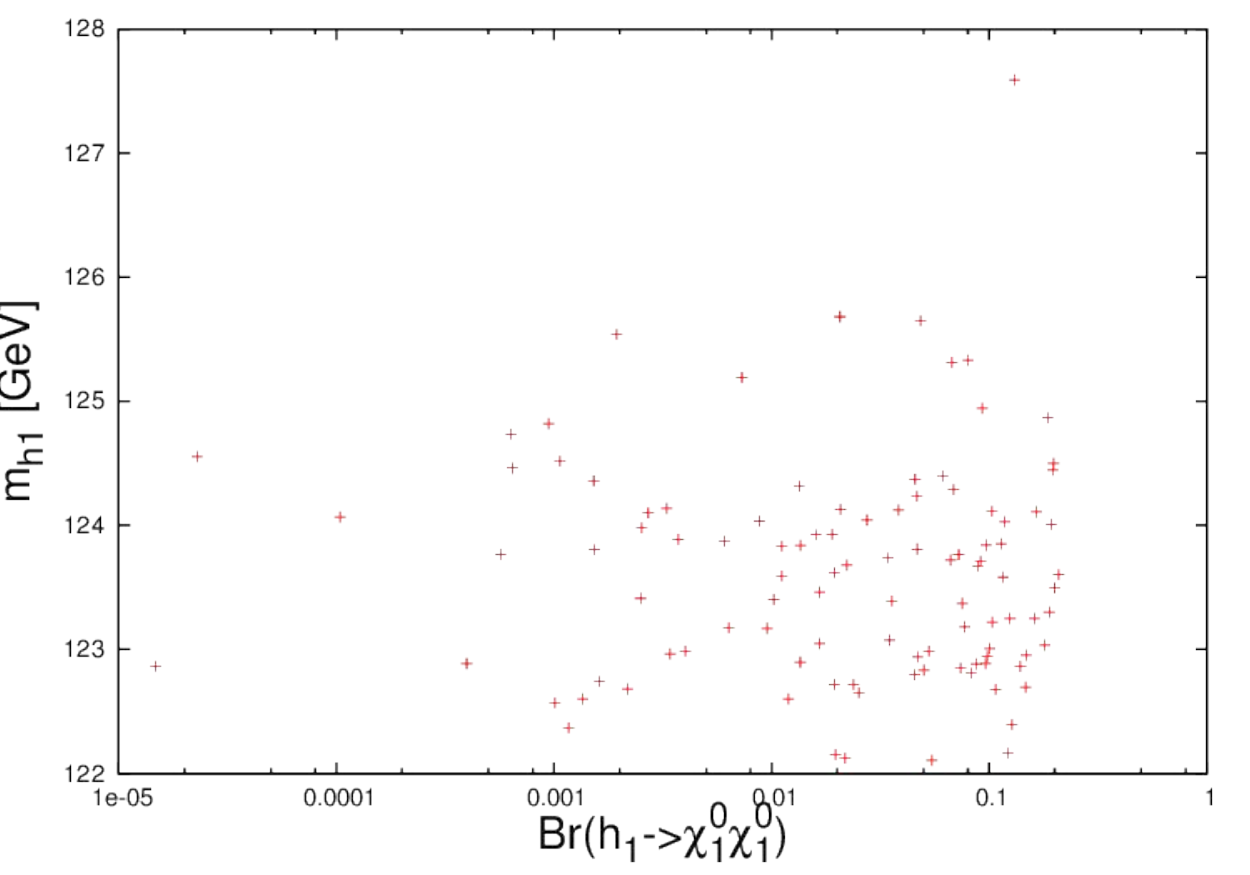}
 \end{tabular}
 \label{fig:mh1-Br}
\caption{The branching ratios Br$(h_1\to b\bar b)$,  Br$(h_1\to \tau^{+}\tau^{-})$, Br$(h_1\to \mu^{+}\mu^{-})$,
Br$(h_1\to \gamma\gamma)$,  Br$(h_1\to Z\gamma)$ and 
Br$(h_1\to \chi^0_1\chi^0_1)$ plotted against the lightest CP-even Higgs mass $m_{h_1}$. }
\end{figure}

The second lightest CP-even Higgs $h_2$ is heavy in the specified parameter regions $m_{h_2}>2m_Z$. Fig. 2 shows the different
patterns of $h_2$ decays. Notice that the branching ratio of $h_2\to b\bar b$ can be dominant even for high values of $m_{h_2}$. 
This is also correct for the decay $a_1\to b\bar b$, see Fig. 3 \footnote{The mass region of $a_1$ below 
the $b\bar b$ threshold is severely 
constrained, see, e.g., Ref. \cite{Lebedev}.}. The dominance of Higgs decay to $b\bar b$ occurs 
when Higgs couplings to down-type fermions are enhanced. Furthermore, it is clear from the Figs. 2 and 3 that
the branching fractions of $h_2\rightarrow\chi^+_1\chi^-_1$, $a_1\rightarrow\gamma\gamma$,  
$a_1\rightarrow Z\gamma$, $a_1\rightarrow\chi^0_1\chi^0_1$ and $a_1\rightarrow\chi^+_1\chi^-_1$ can be dominant in some area of
the selected parameter space when the the Higgs boson is  highly singlet.
\newpage
\begin{figure}
 \centering\begin{tabular}{ccc}
 \includegraphics[scale=0.450]{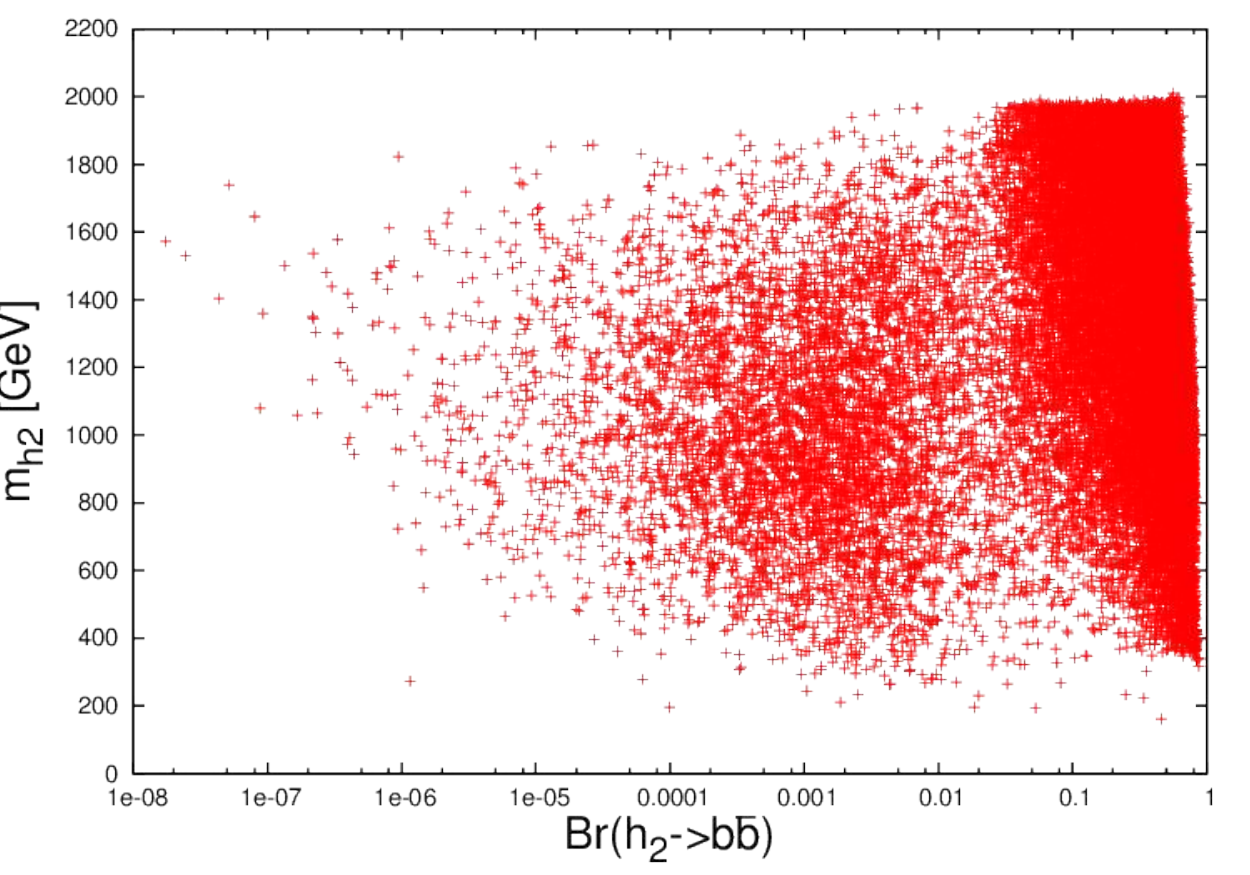}
 &\includegraphics[scale=0.450]{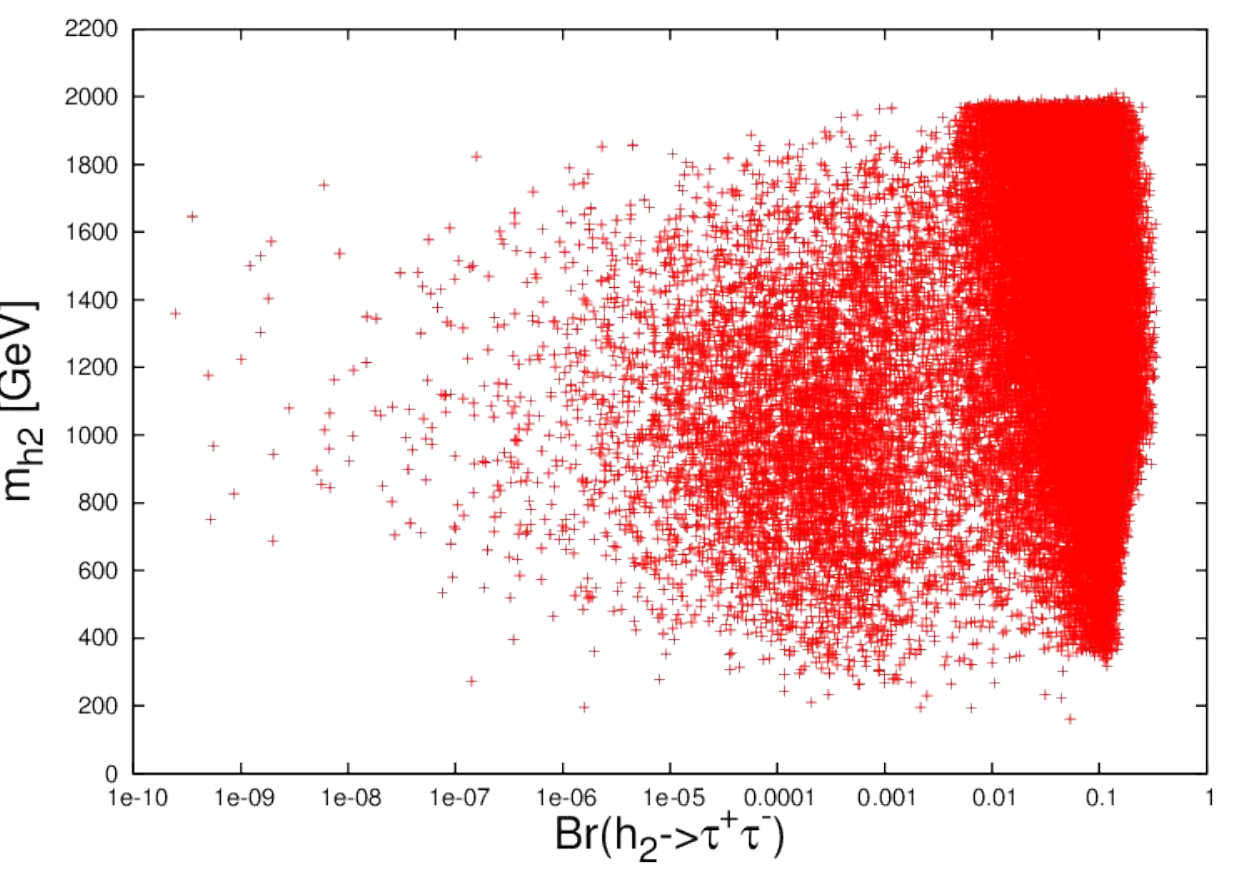} \\
 \includegraphics[scale=0.450]{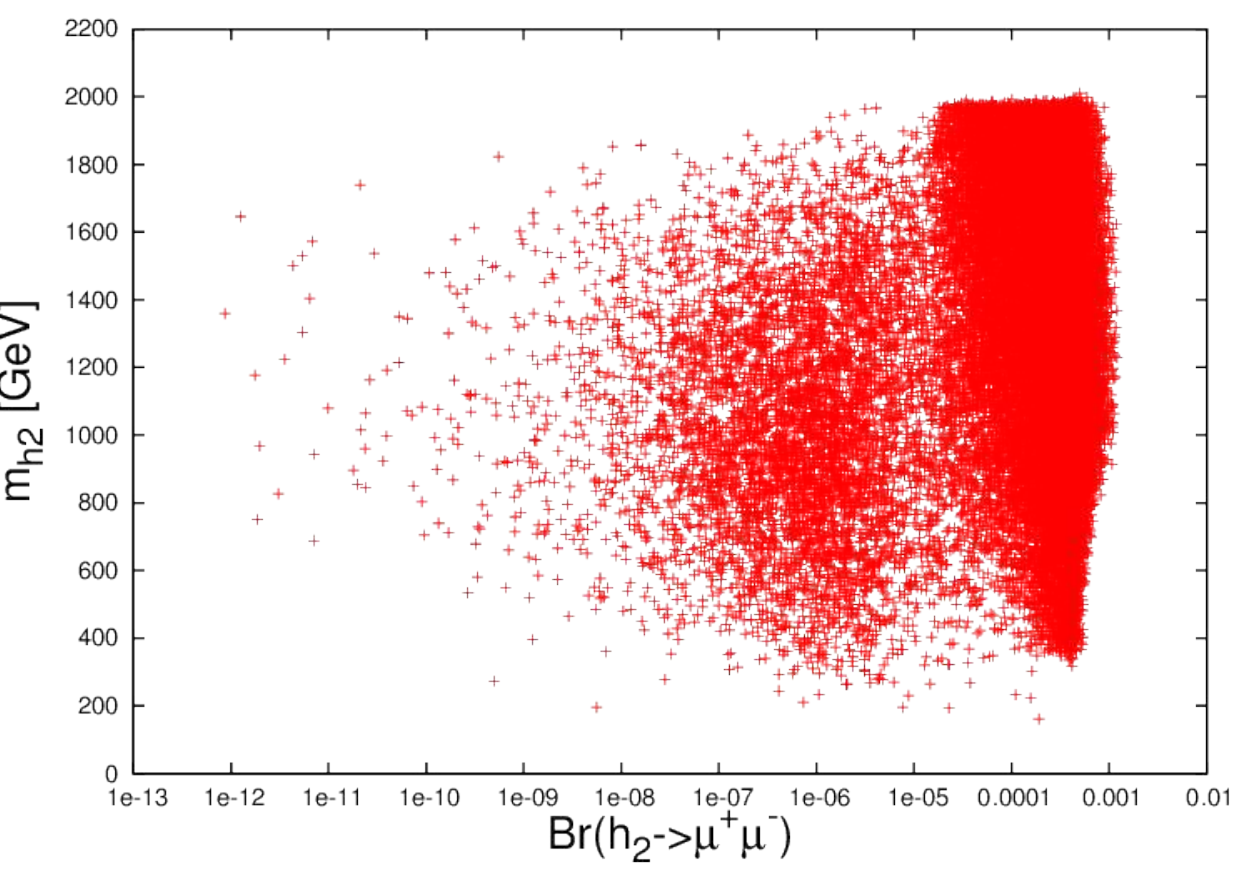} 
 &\includegraphics[scale=0.450]{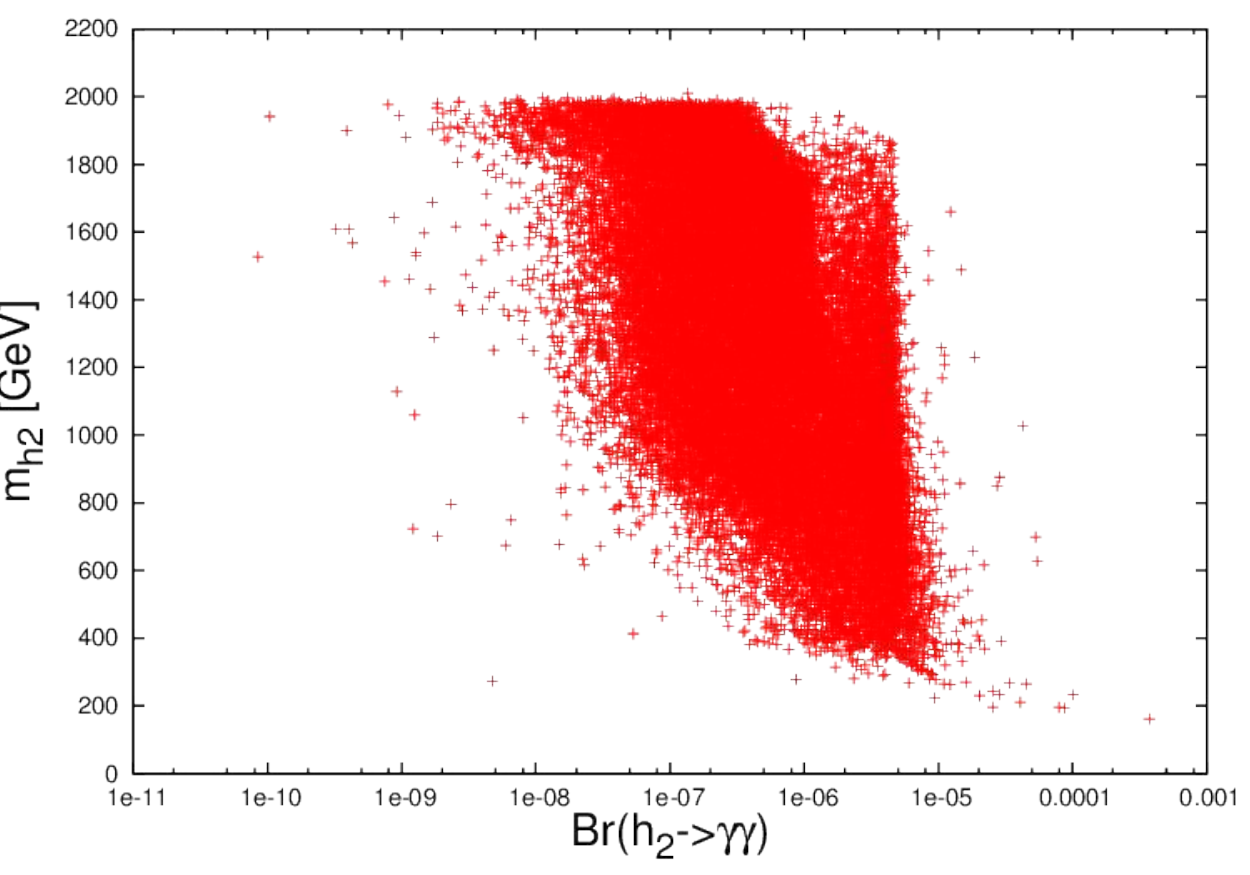}\\
 \includegraphics[scale=0.450]{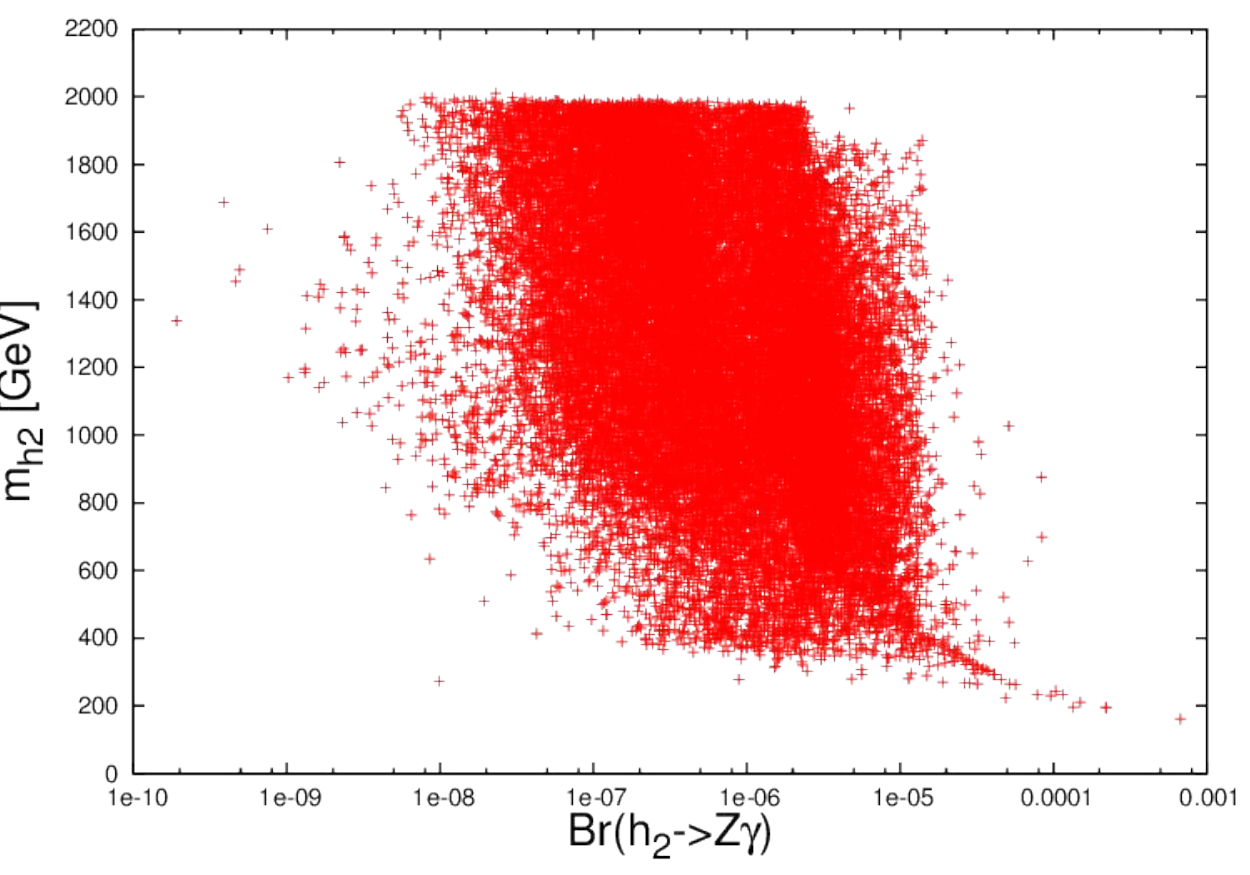}
 &\includegraphics[scale=0.450]{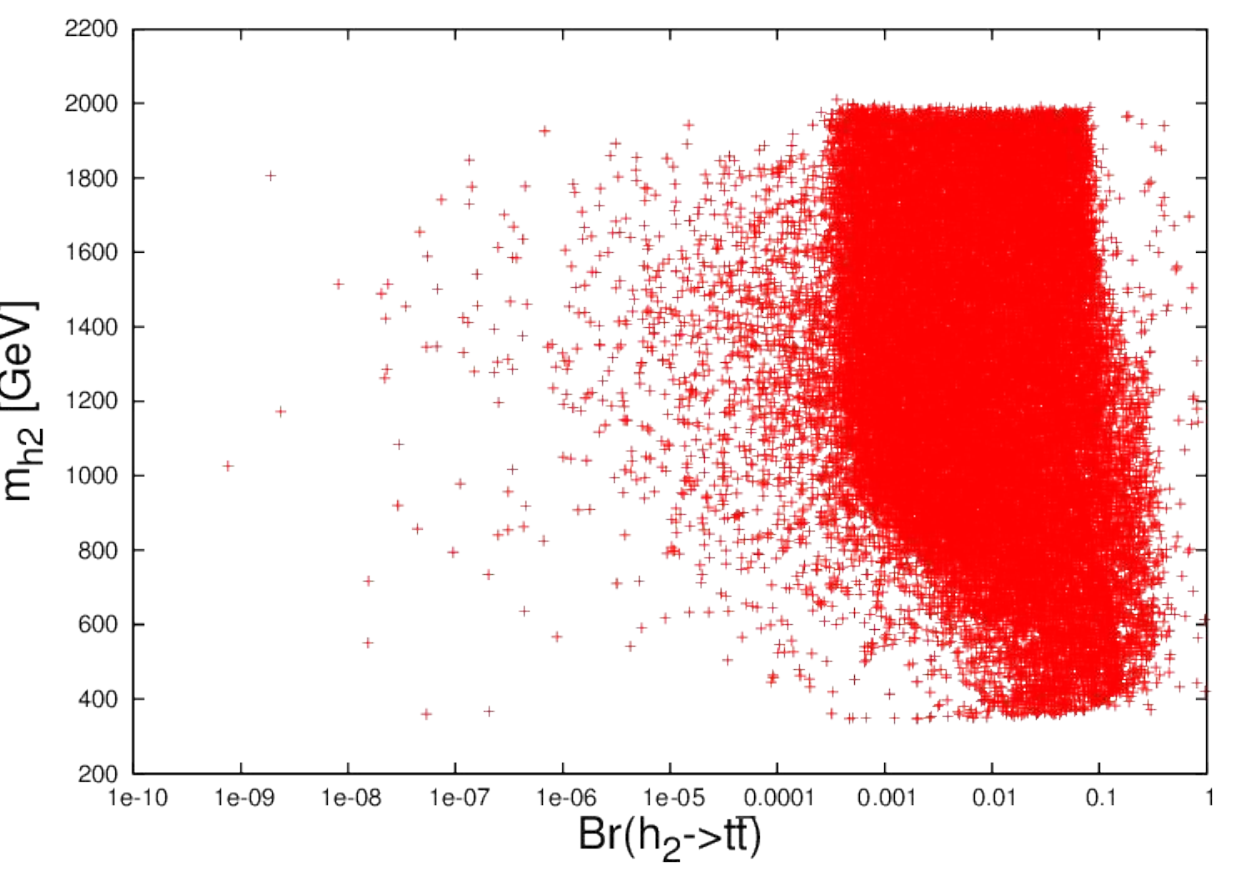}\\
 \includegraphics[scale=0.450]{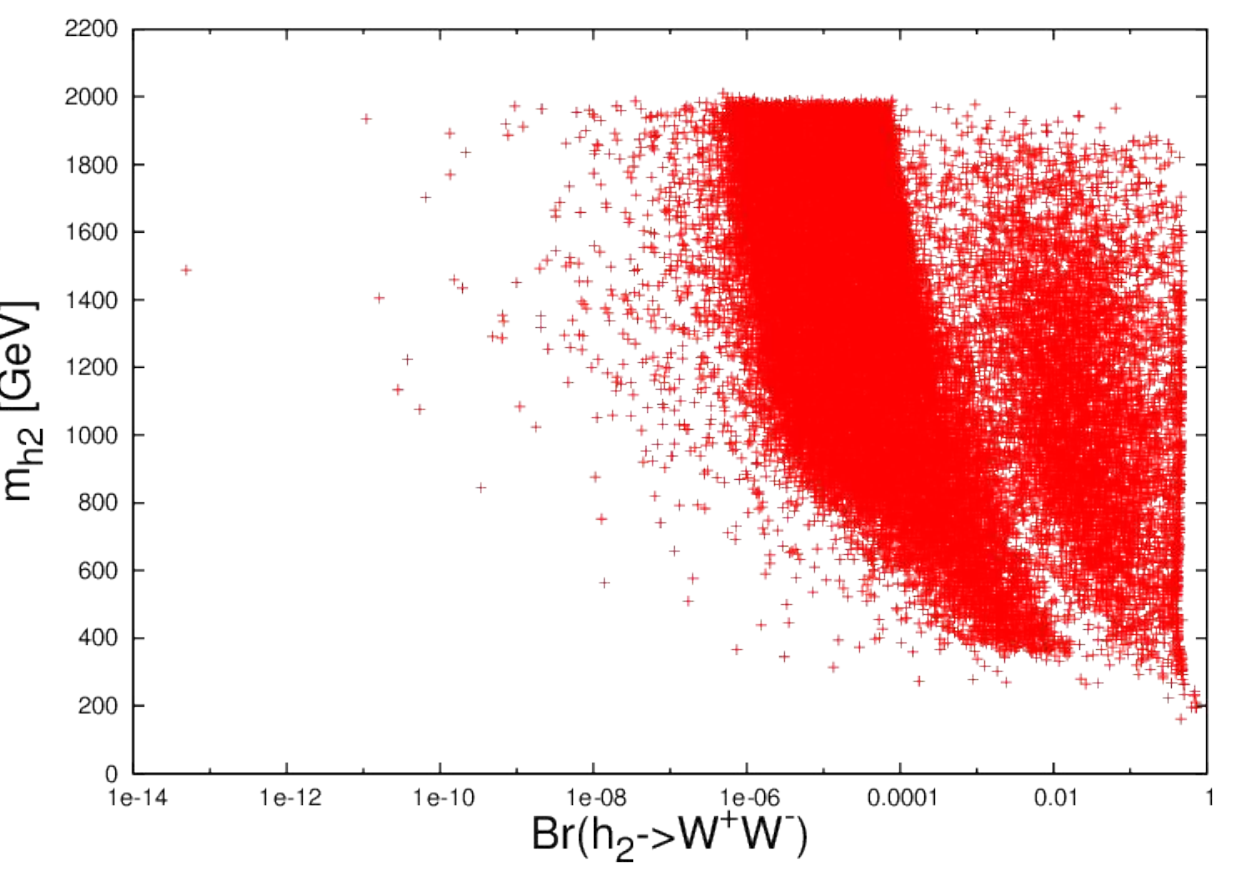}
 &\includegraphics[scale=0.450]{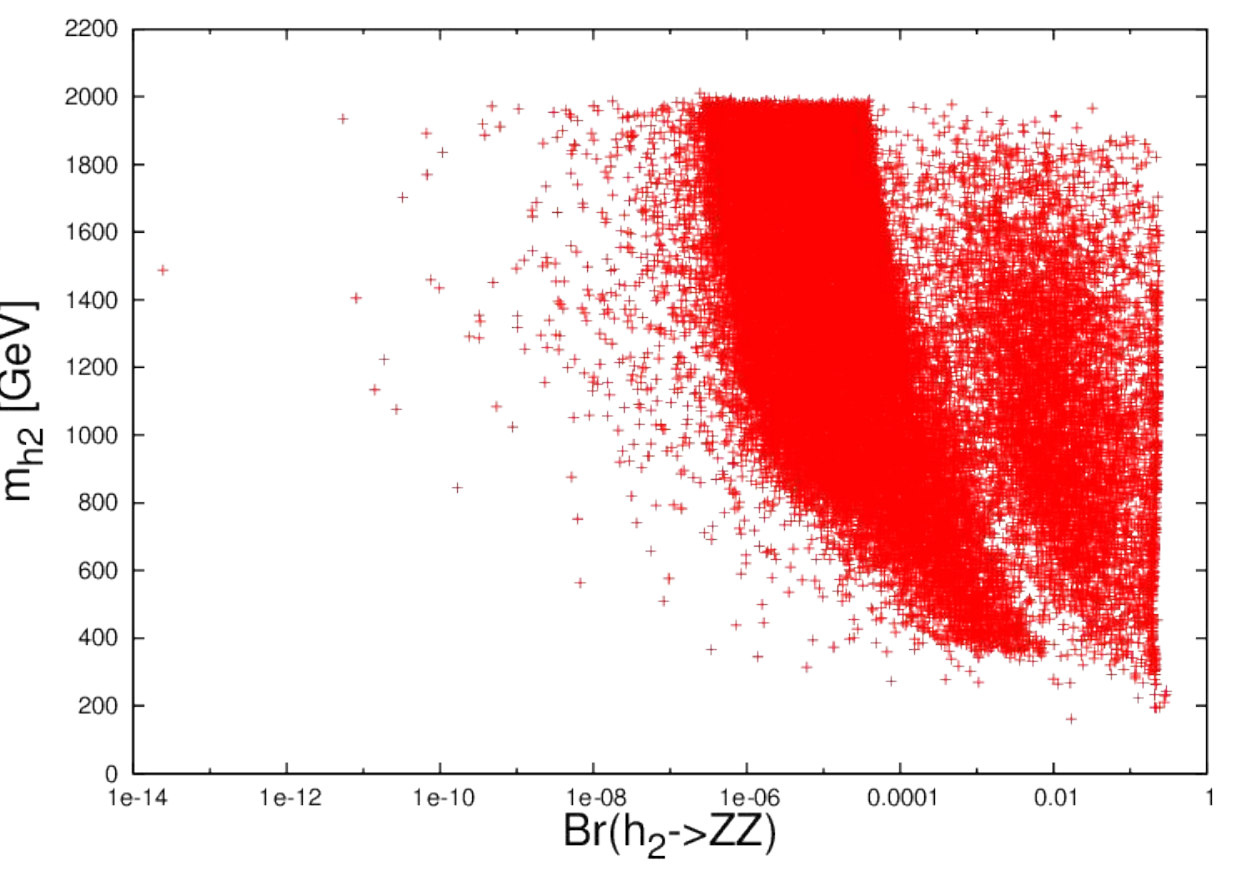} \\
 \includegraphics[scale=0.450]{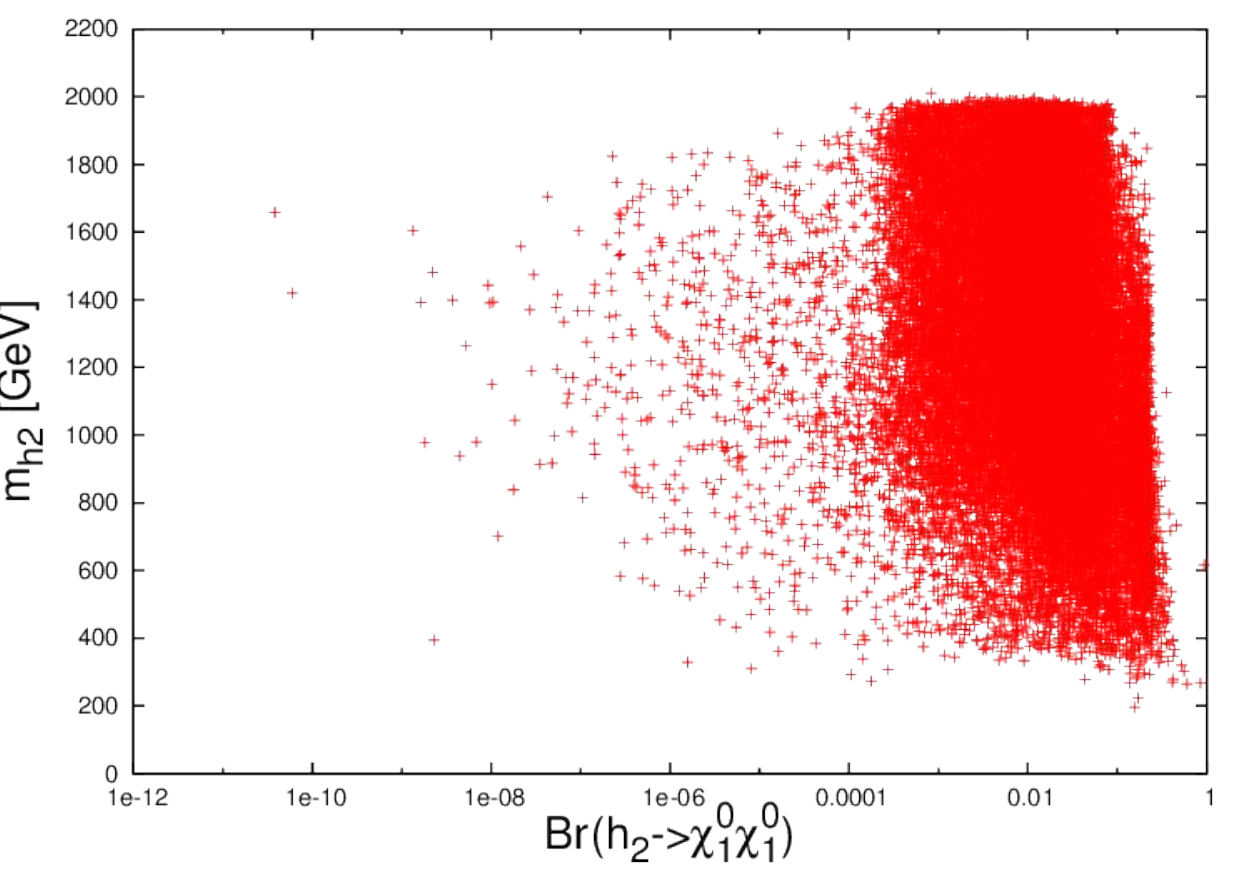}
& \includegraphics[scale=0.450]{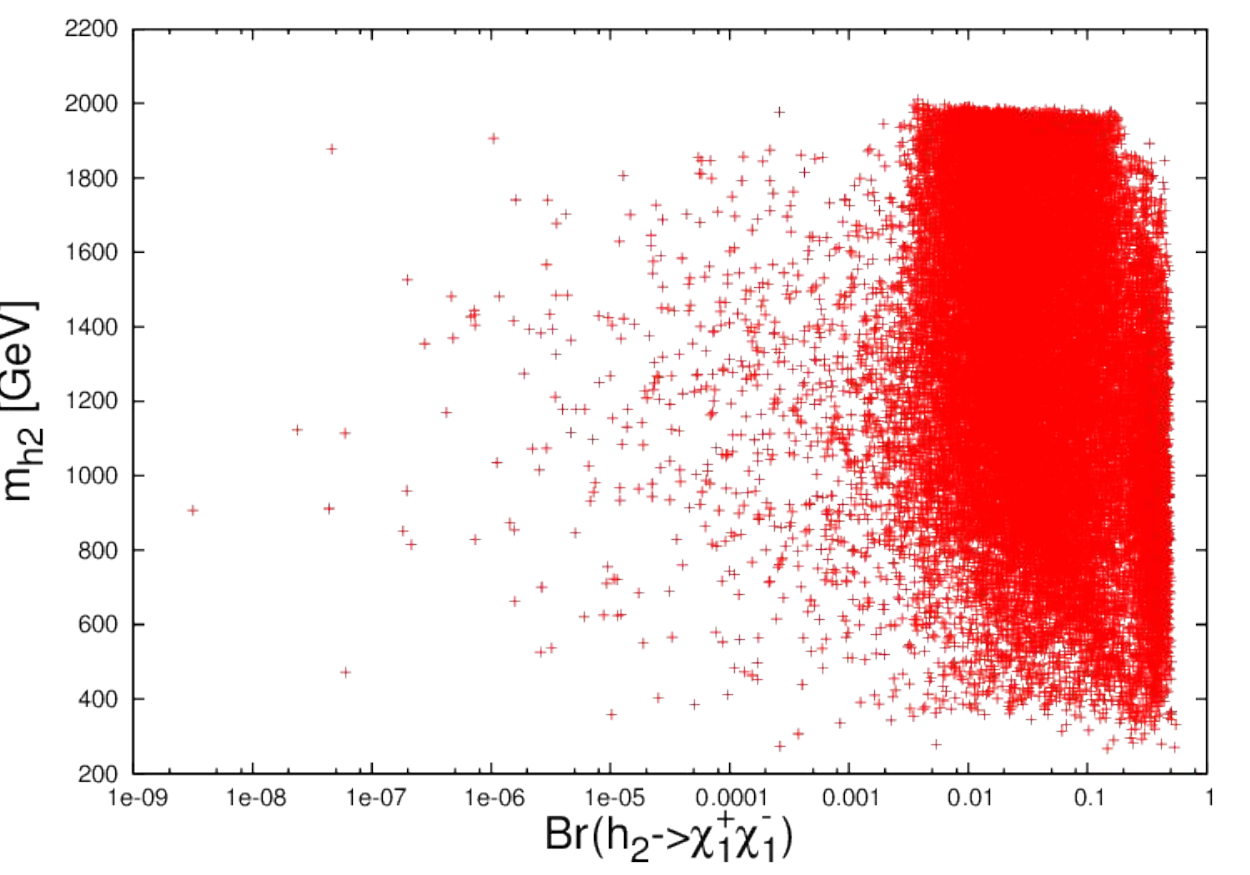}
 \end{tabular}
 \label{fig:mh2-Br}
\caption{The branching ratios Br$(h_2\to b\bar b)$,  Br$(h_2\to \tau^{+}\tau^{-})$, Br$(h_2\to \mu^{+}\mu^{-})$, 
Br$(h_2\to \gamma\gamma)$, Br$(h_2\to Z\gamma)$,
Br$(h_2\to t\bar t)$, Br$(h_2\to W^{+}W^{-})$, Br$(h_2\to ZZ)$, 
Br$(h_2\to \chi^0_1\chi^0_1)$ and Br$(h_2\to \chi^+_1\chi^-_1)$ plotted against the second lightest CP-even Higgs mass $m_{h_2}$. }
\end{figure}

\newpage
\begin{figure}
 \centering\begin{tabular}{ccc}
 \includegraphics[scale=0.50]{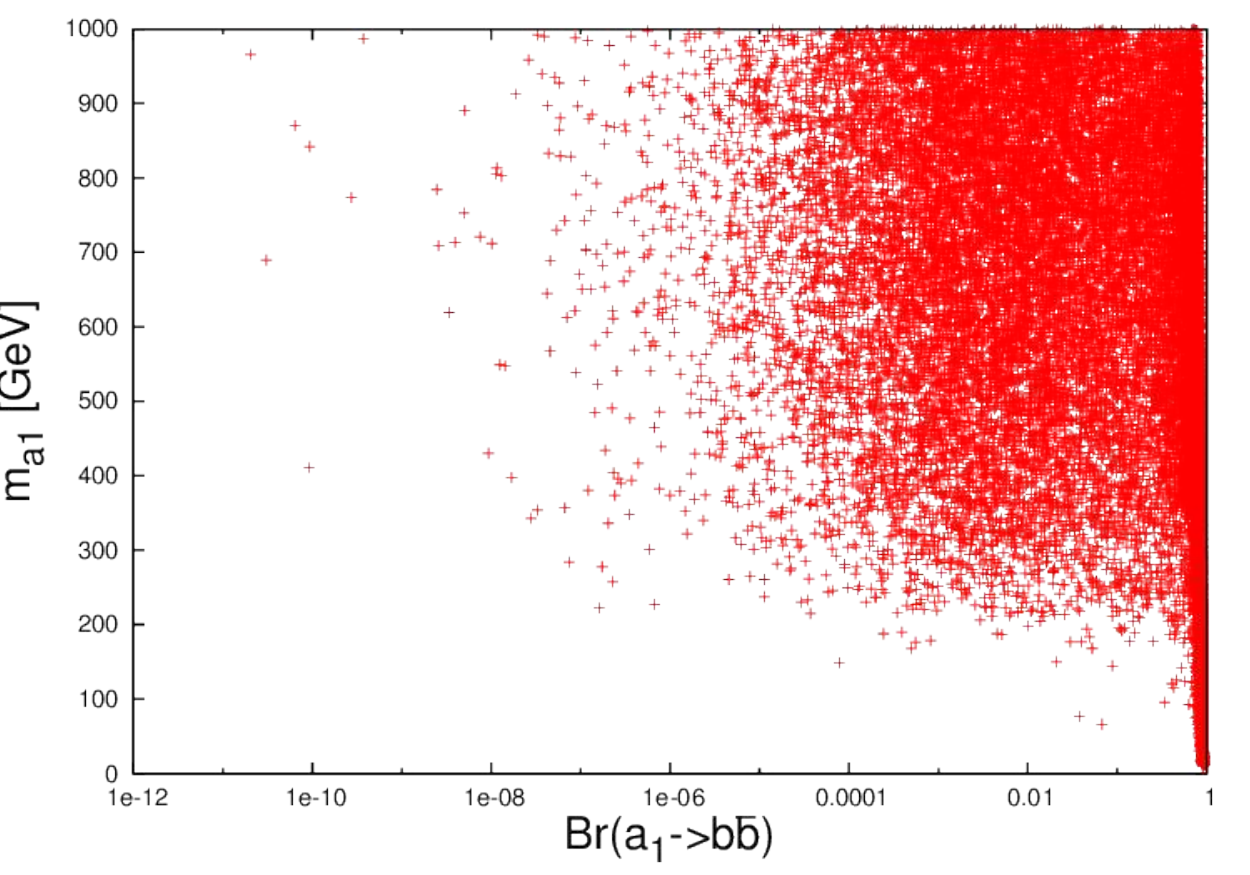}
 &\includegraphics[scale=0.50]{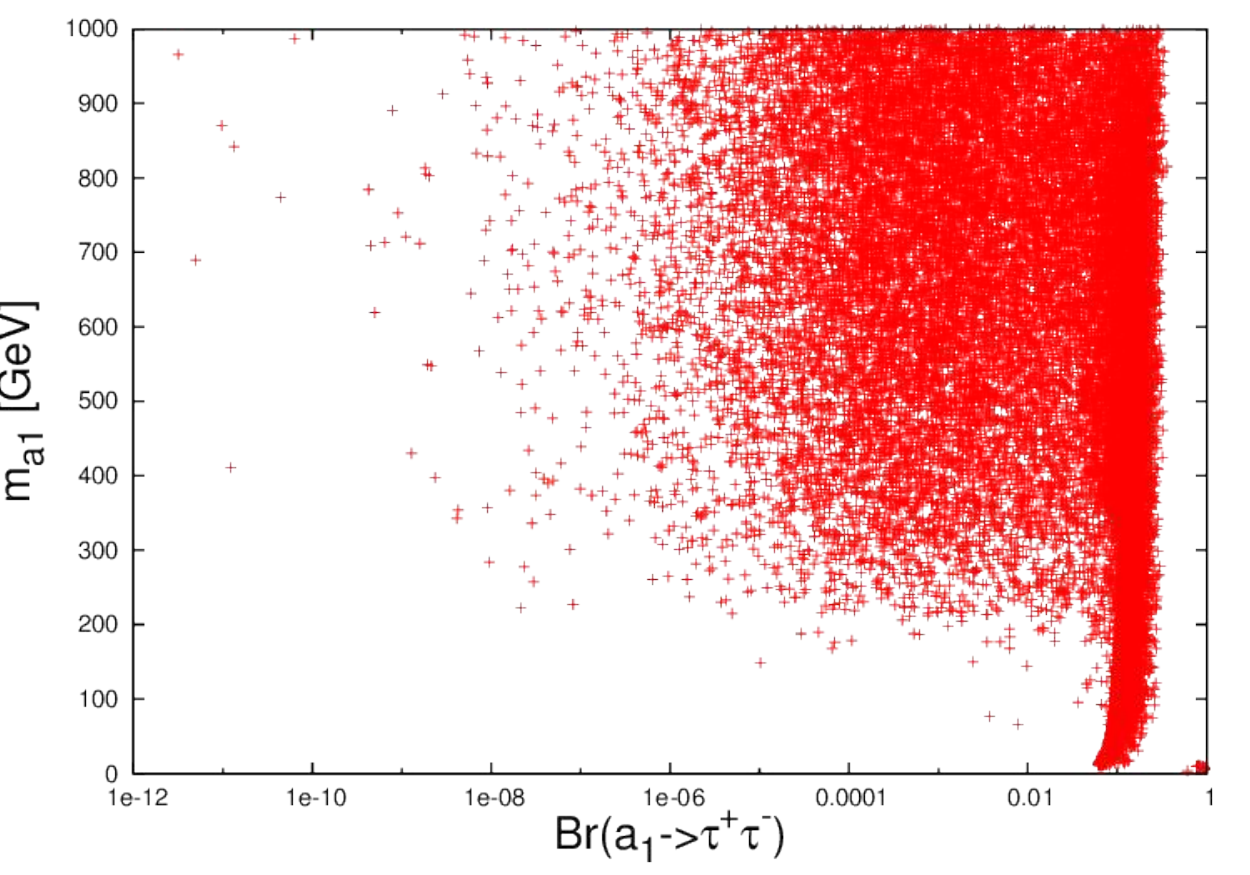} \\
 \includegraphics[scale=0.50]{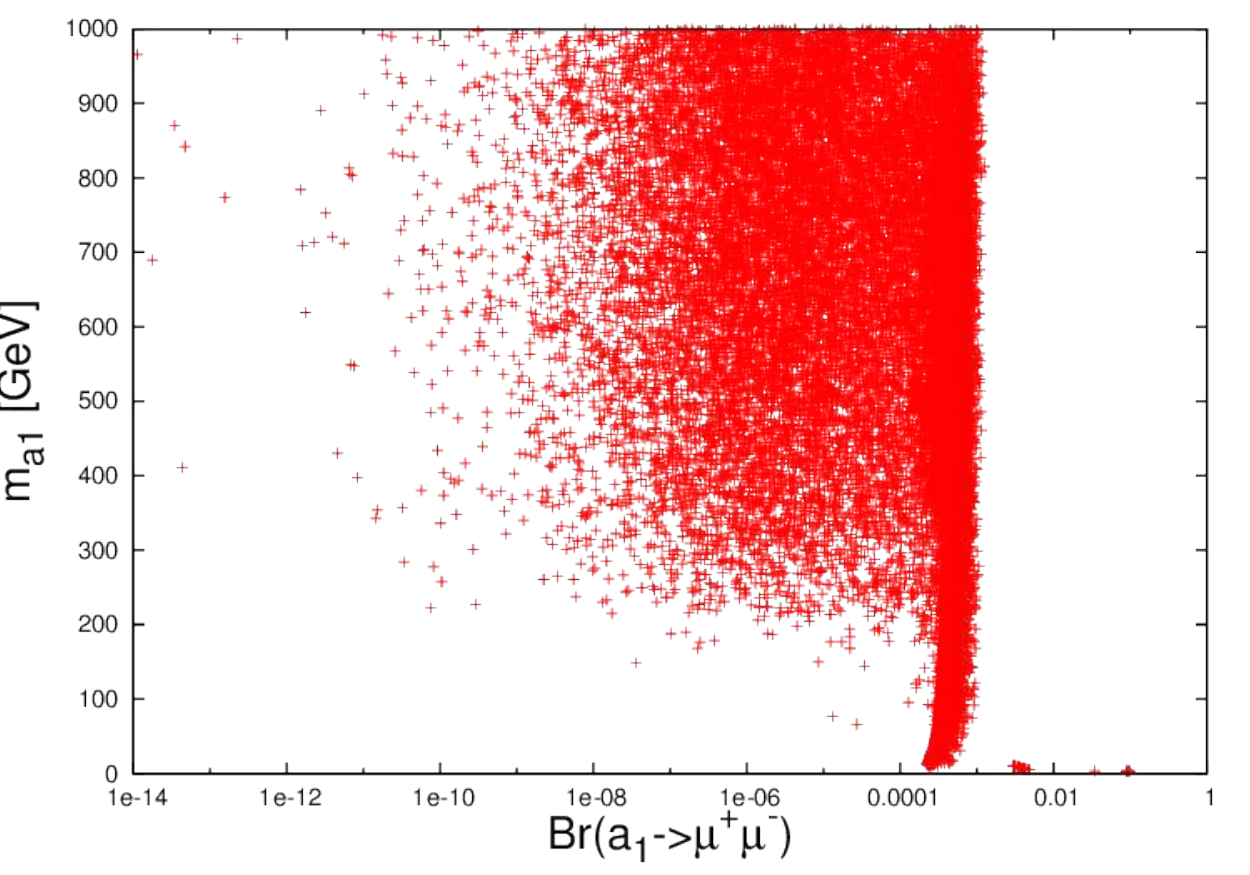}
 &\includegraphics[scale=0.50]{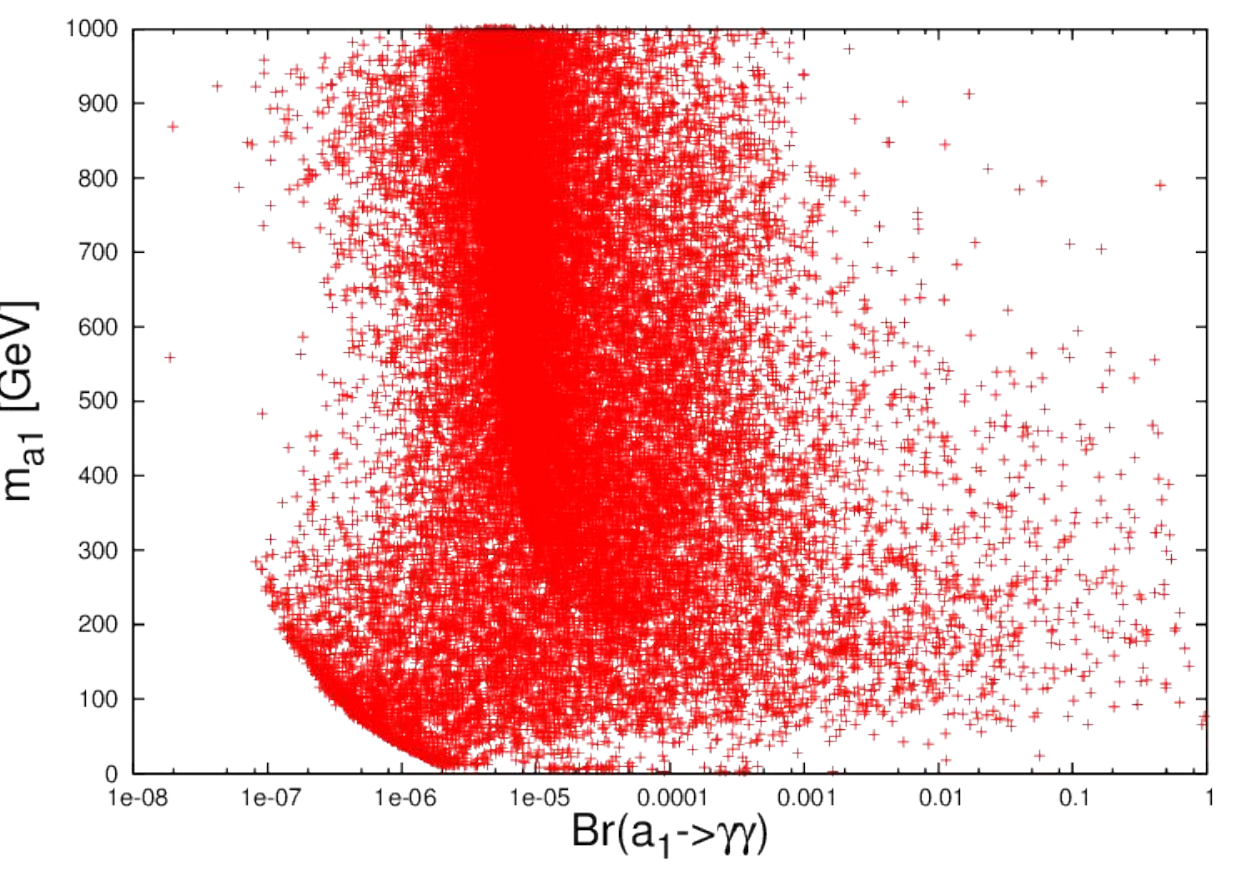} \\
 \includegraphics[scale=0.50]{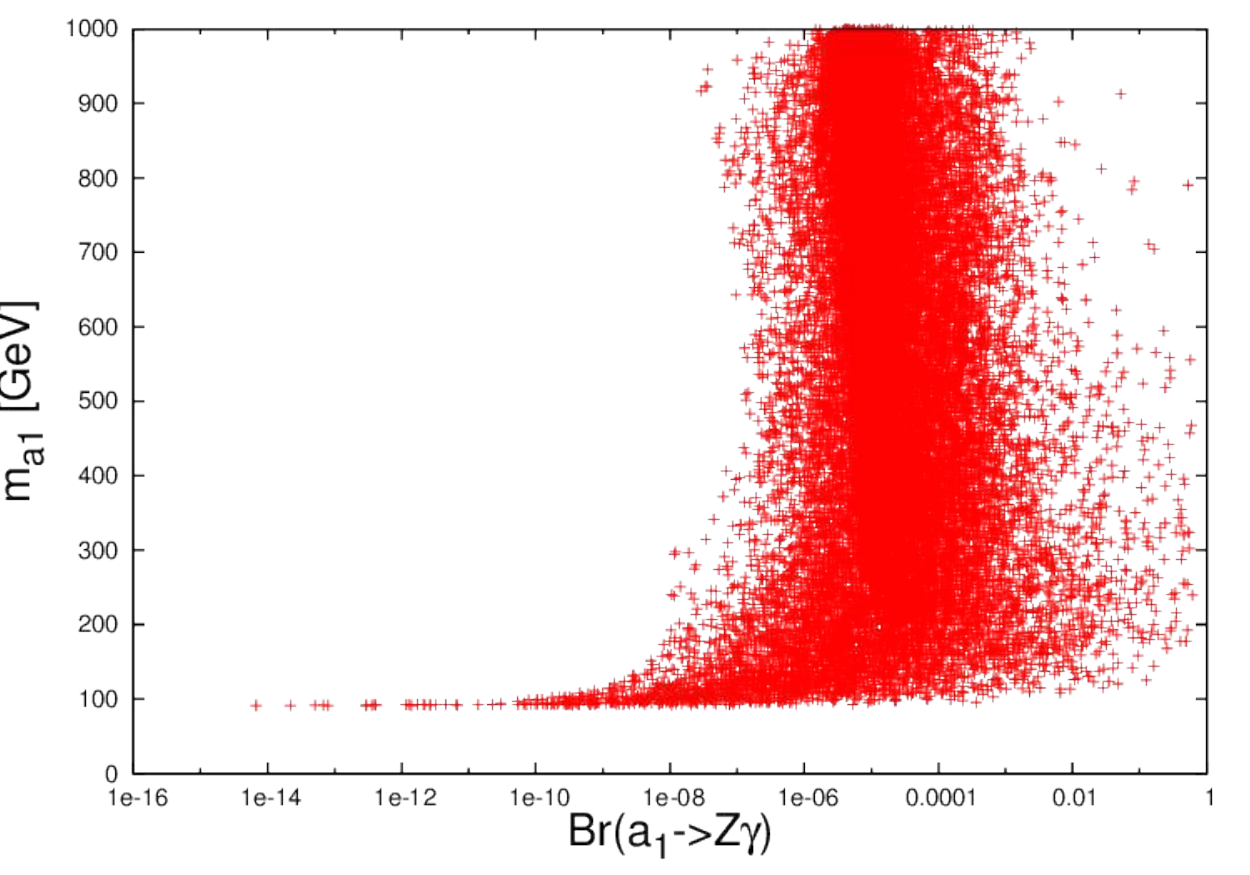}
 &\includegraphics[scale=0.50]{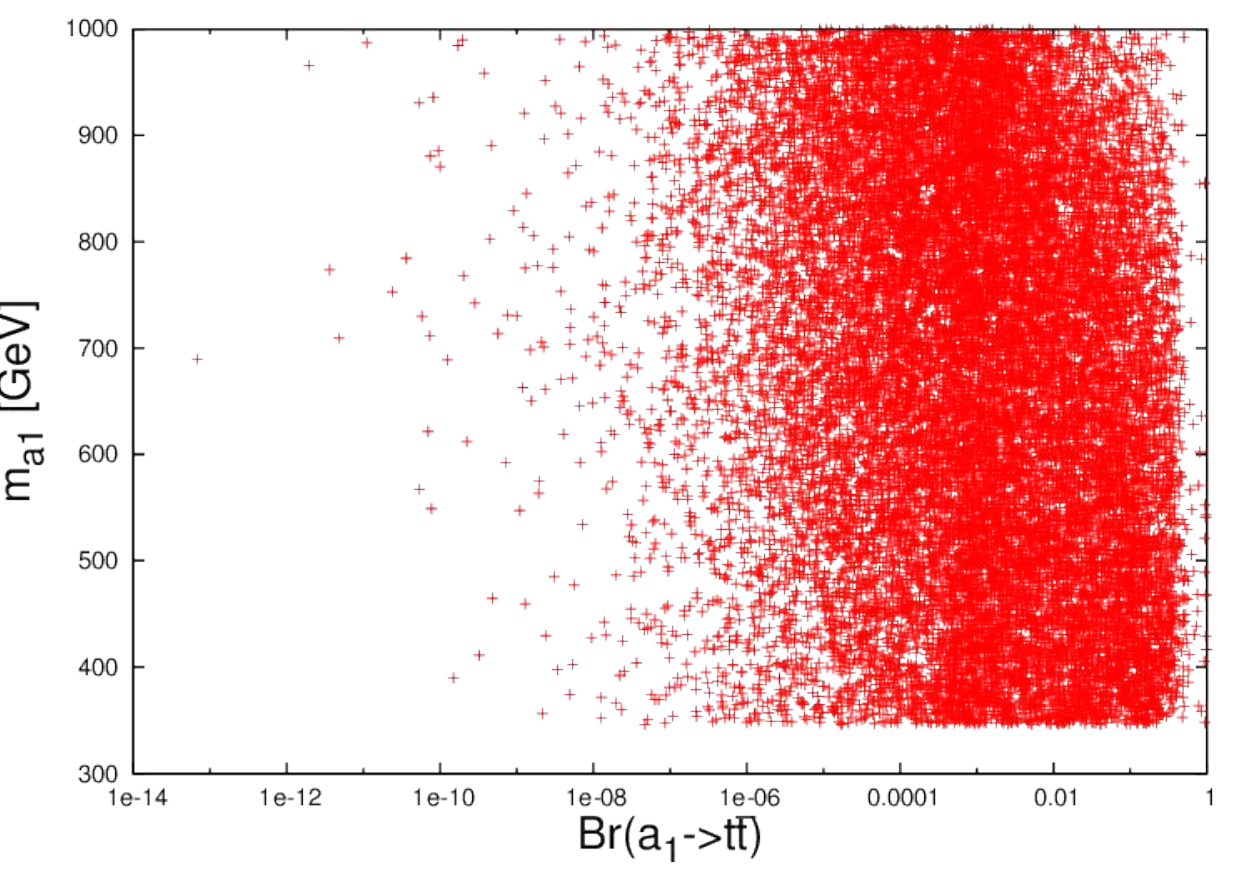}\\
 \includegraphics[scale=0.50]{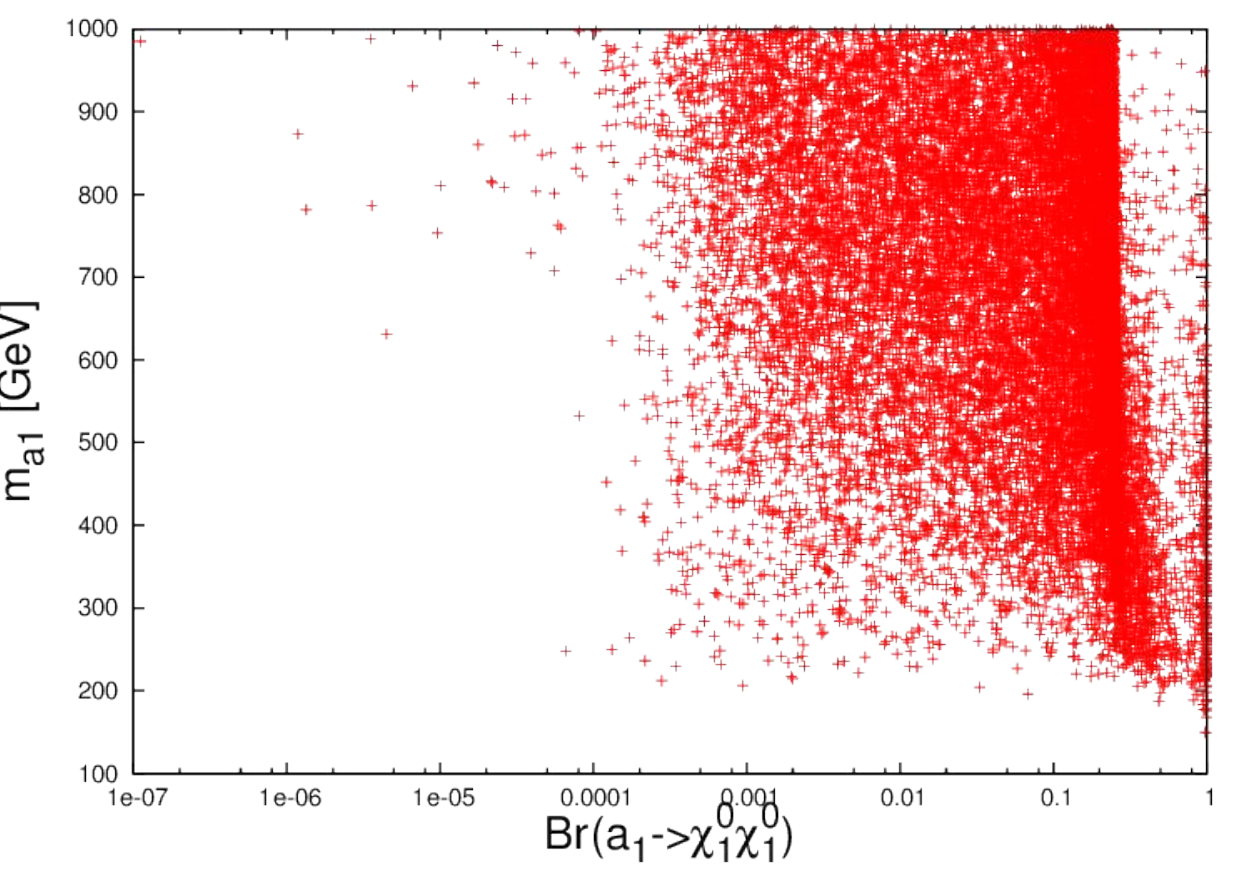}
 &\includegraphics[scale=0.50]{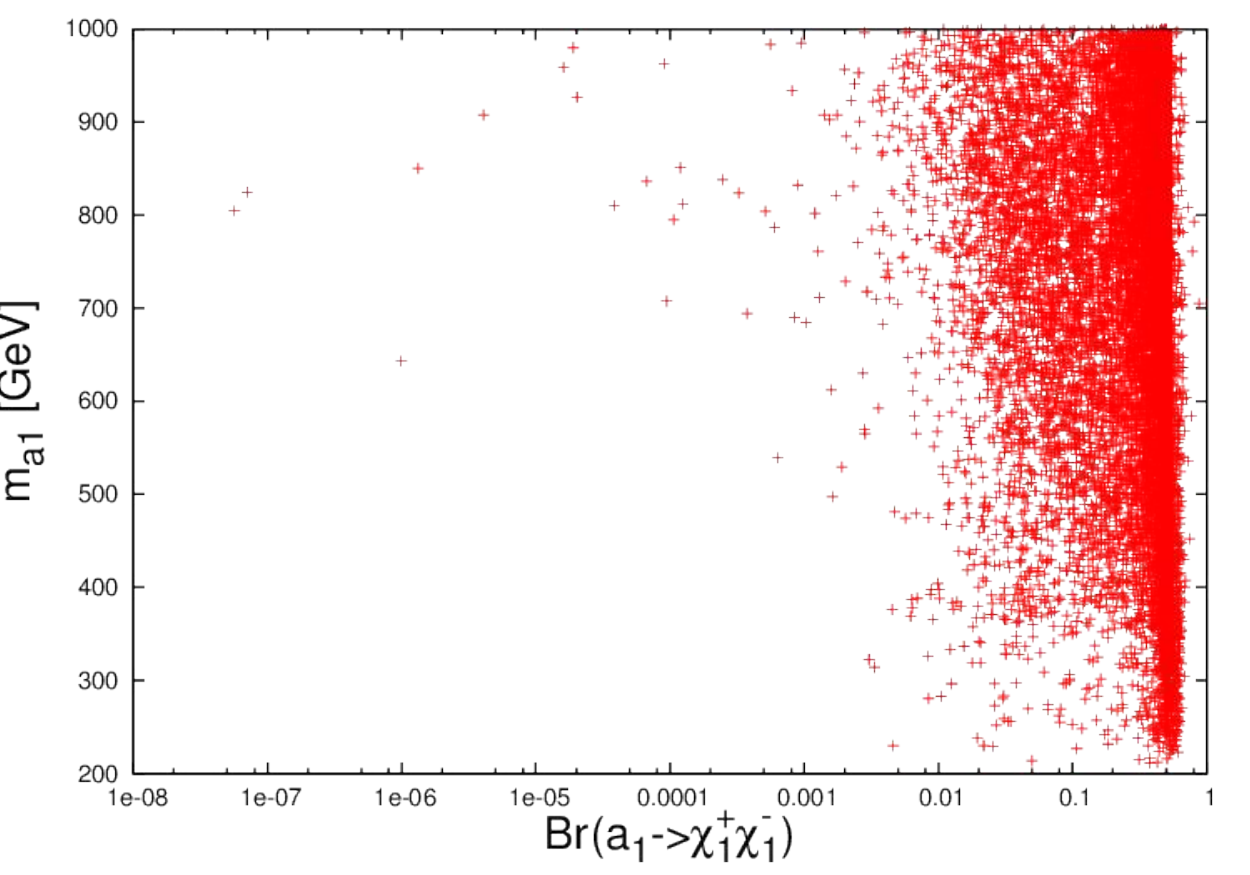}
\end{tabular}
\label{fig:ma1-Br}
\caption{The branching ratios Br$(a_1\to b\bar b)$, Br$(a_1\to \tau^{+}\tau^{-})$, Br$(a_1\to \mu^{+}\mu^{-})$,
Br$(a_1\to \gamma\gamma)$, Br$(a_1\to Z\gamma)$, Br$(a_1\to t\bar t)$, 
Br$(a_1\to \chi^0_1\chi^0_1)$ and Br$(a_1\to \chi^+_1\chi^-_1)$ plotted against the lightest CP-odd Higgs mass $m_{a_1}$. }

\end{figure}

The correlations between all three Higgs masses $m_{h_1}$, $m_{h_2}$ and $m_{a_1}$, and various Higgs-Higgs decays
($h_{1, 2}\to a_1a_1$, $h_2\to h_1h_1$ and $a_1\to Zh_{1, 2}$) are shown in Fig 4. It is clear that the Higgs-to-Higgs decays should be
taken very seriously when searching for $h_2$ at the LHC. This is because there are large regions of the parameter space where
the branching ratios of $h_2\to h_1h_1$ and
$h_2\to a_1a_1$ are sizable. The latter one can reach unity for all $m_{h_2}$. It is also obvious that the branching fractions of
$a_1\to Zh_{1, 2}$ is small, reaching its maximum at around $1 \%$. 
\begin{figure}
 \centering\begin{tabular}{ccc}
 \includegraphics[scale=0.50]{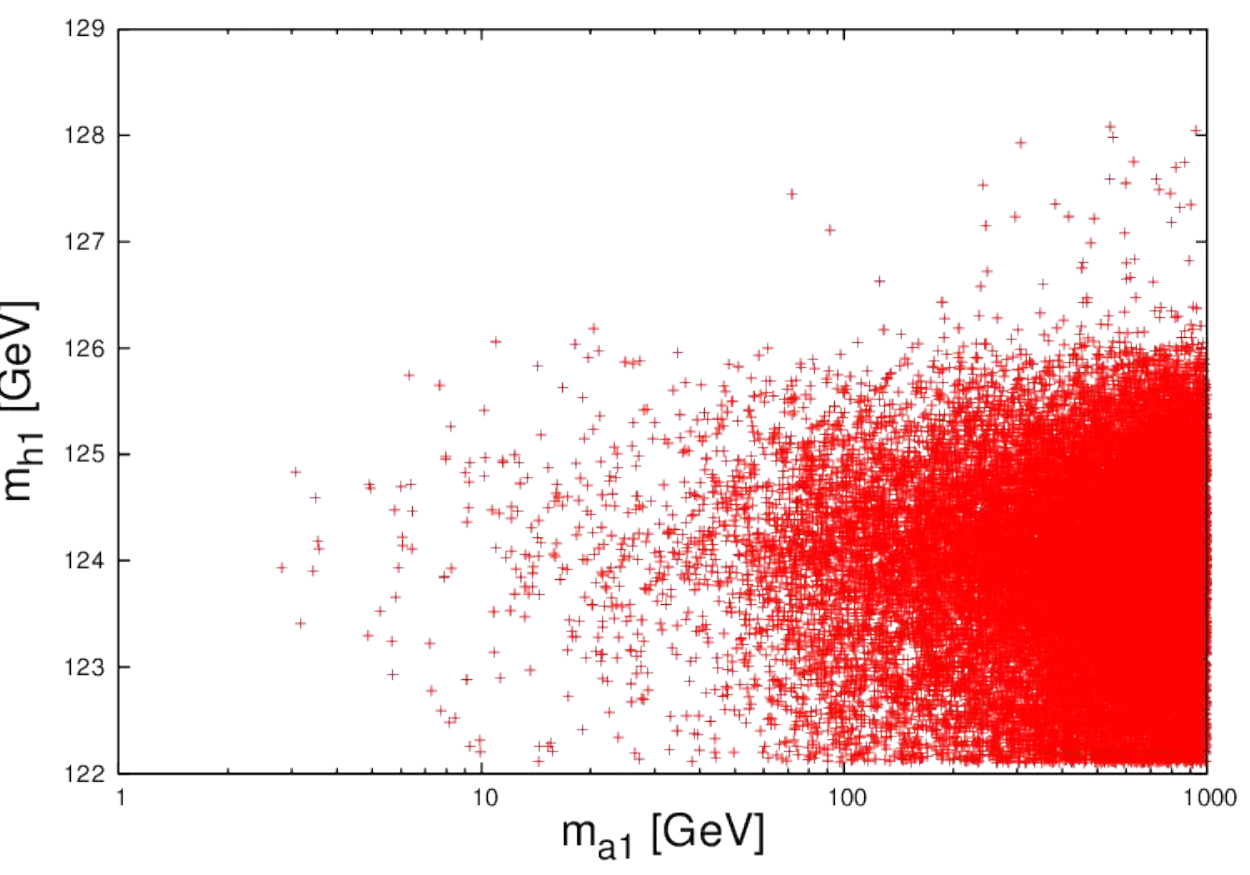}
  &\includegraphics[scale=0.50]{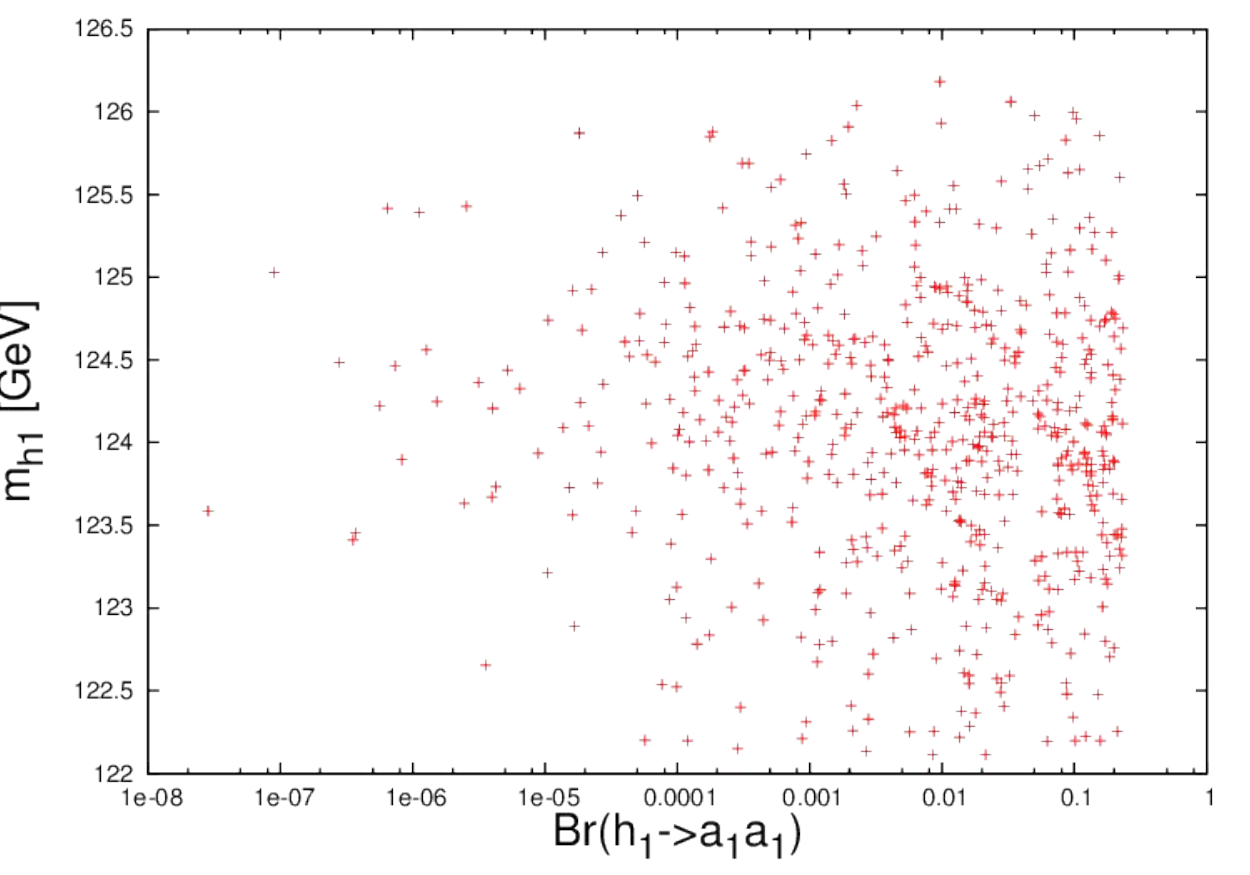} \\
\includegraphics[scale=0.50]{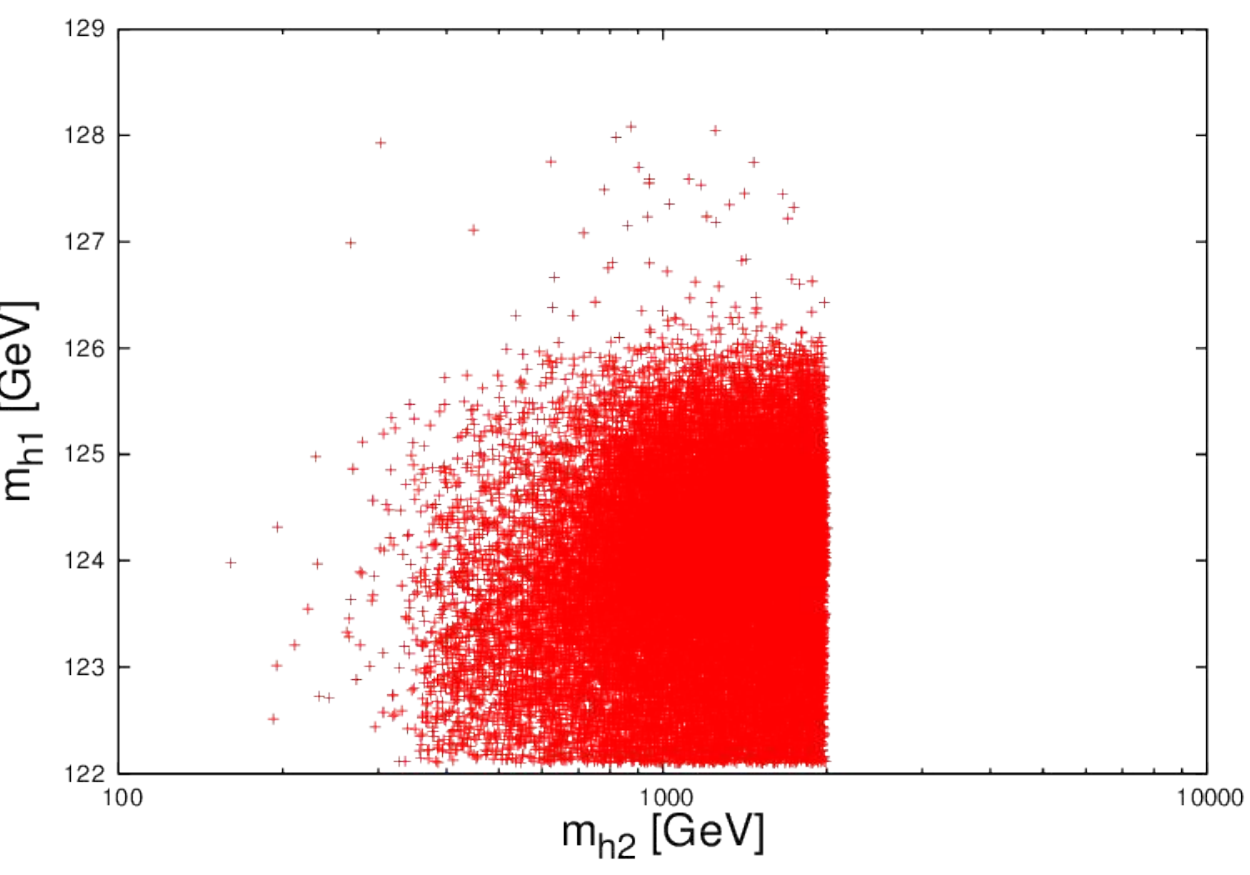}
  &\includegraphics[scale=0.50]{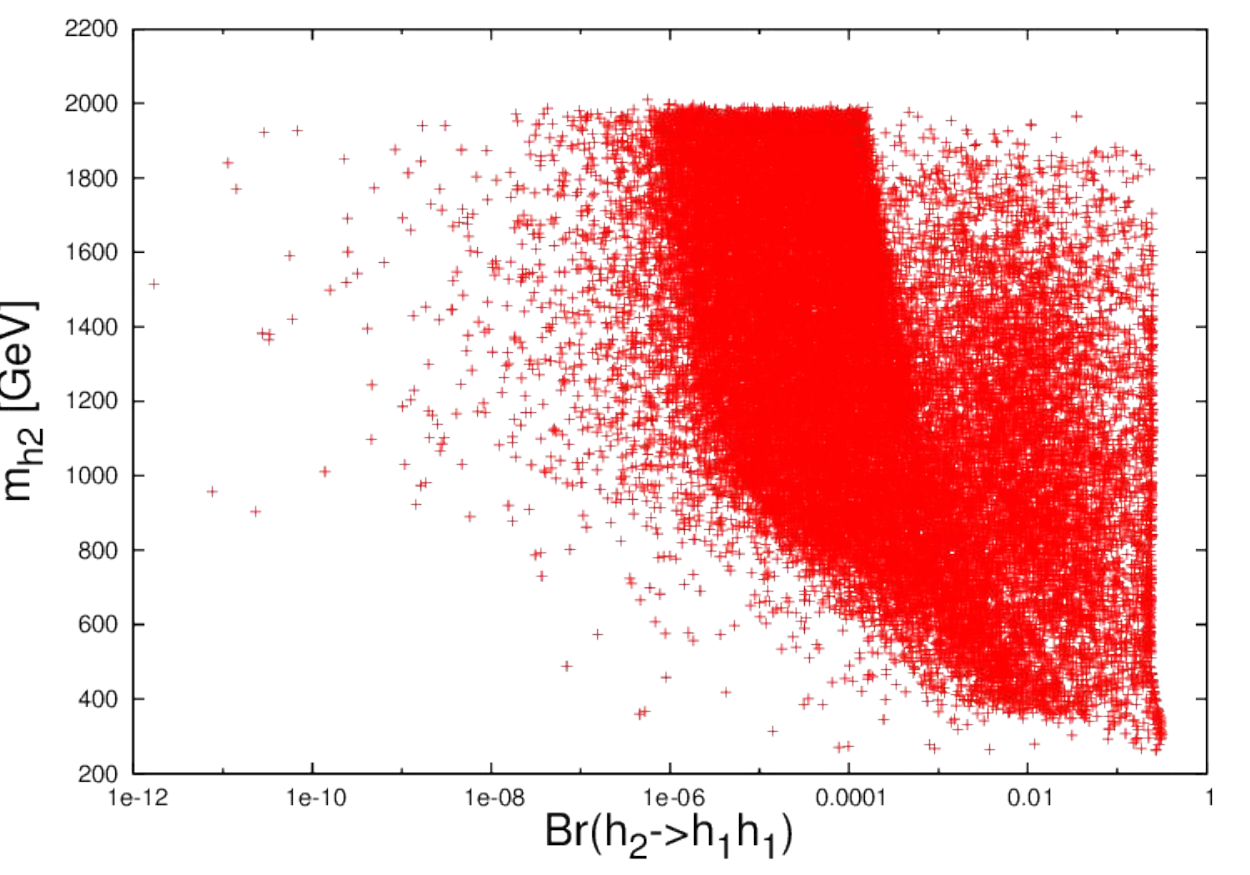} \\
 \includegraphics[scale=0.50]{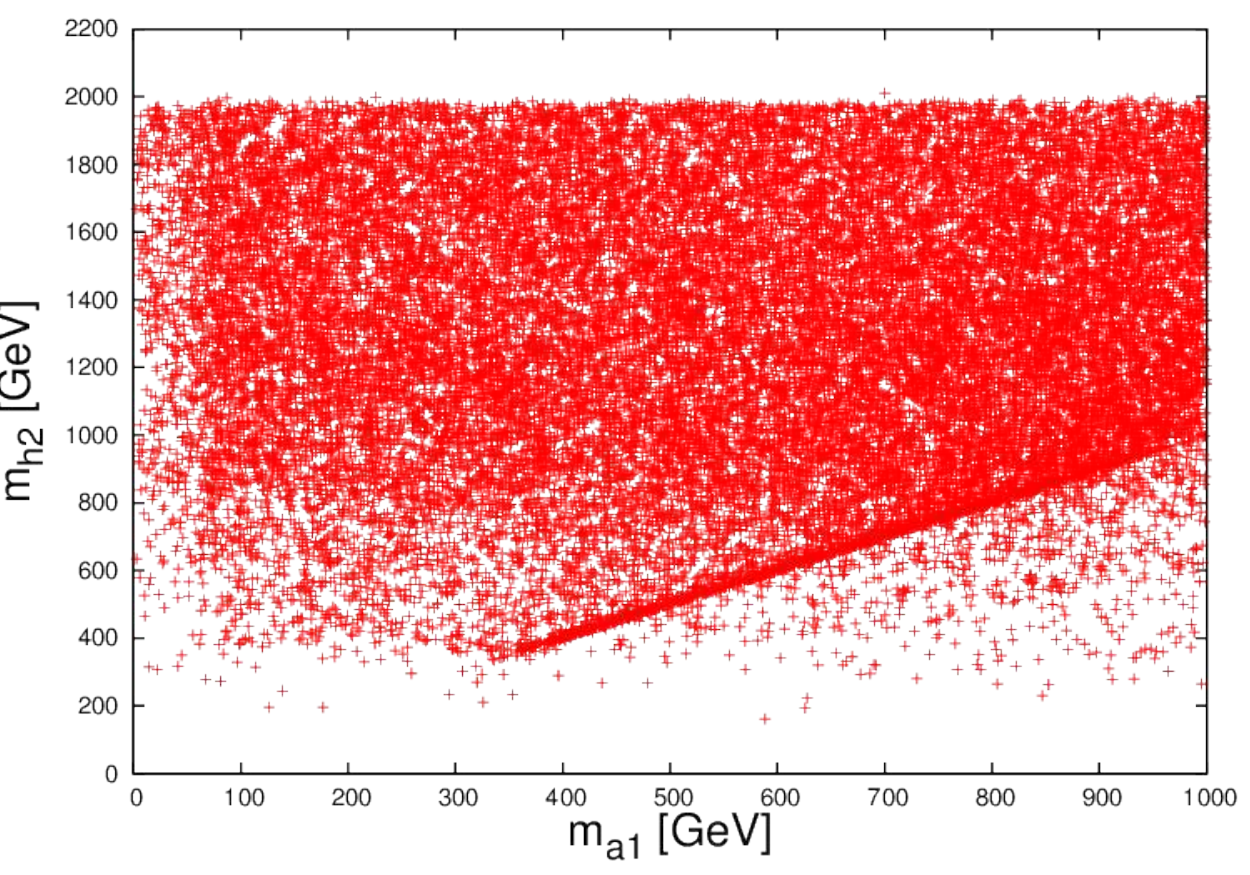}
 &\includegraphics[scale=0.50]{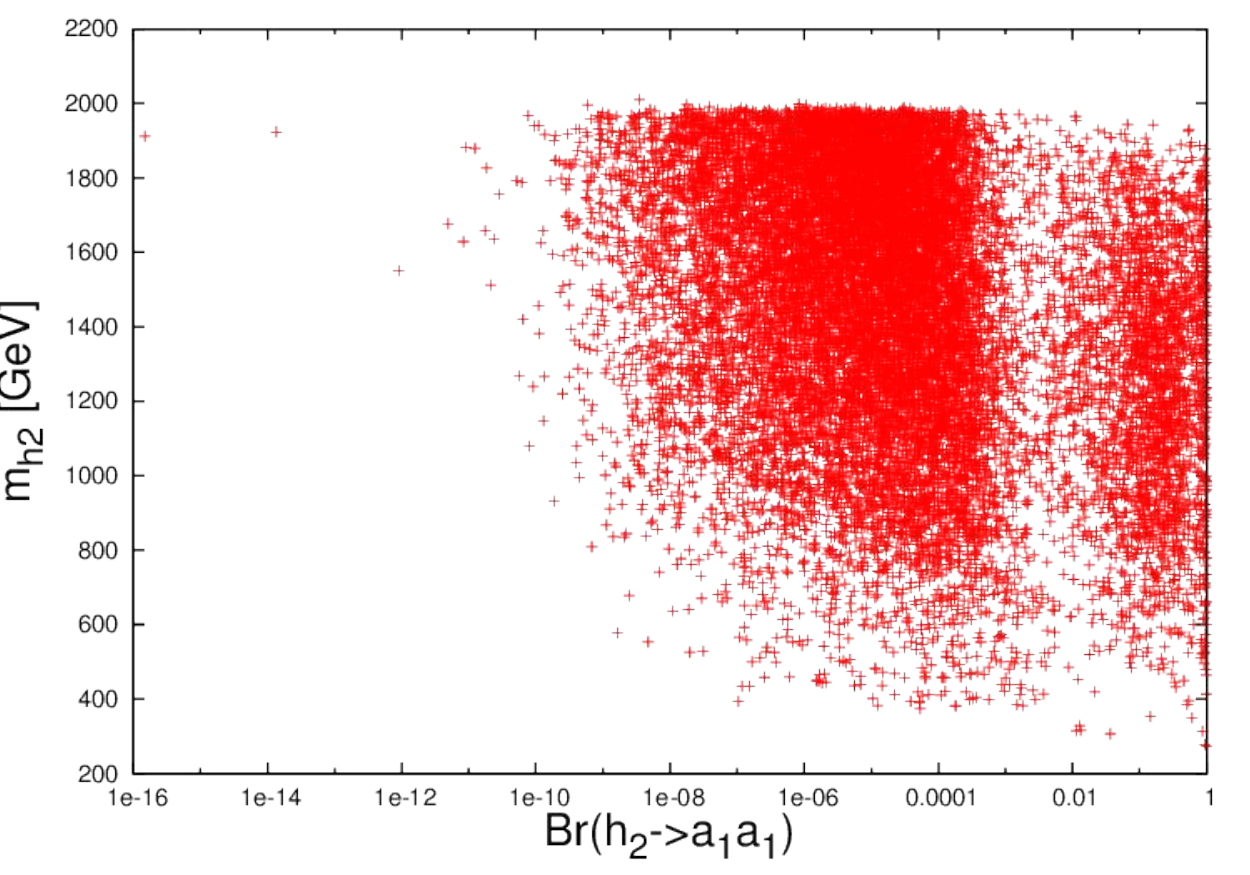}\\
 \includegraphics[scale=0.50]{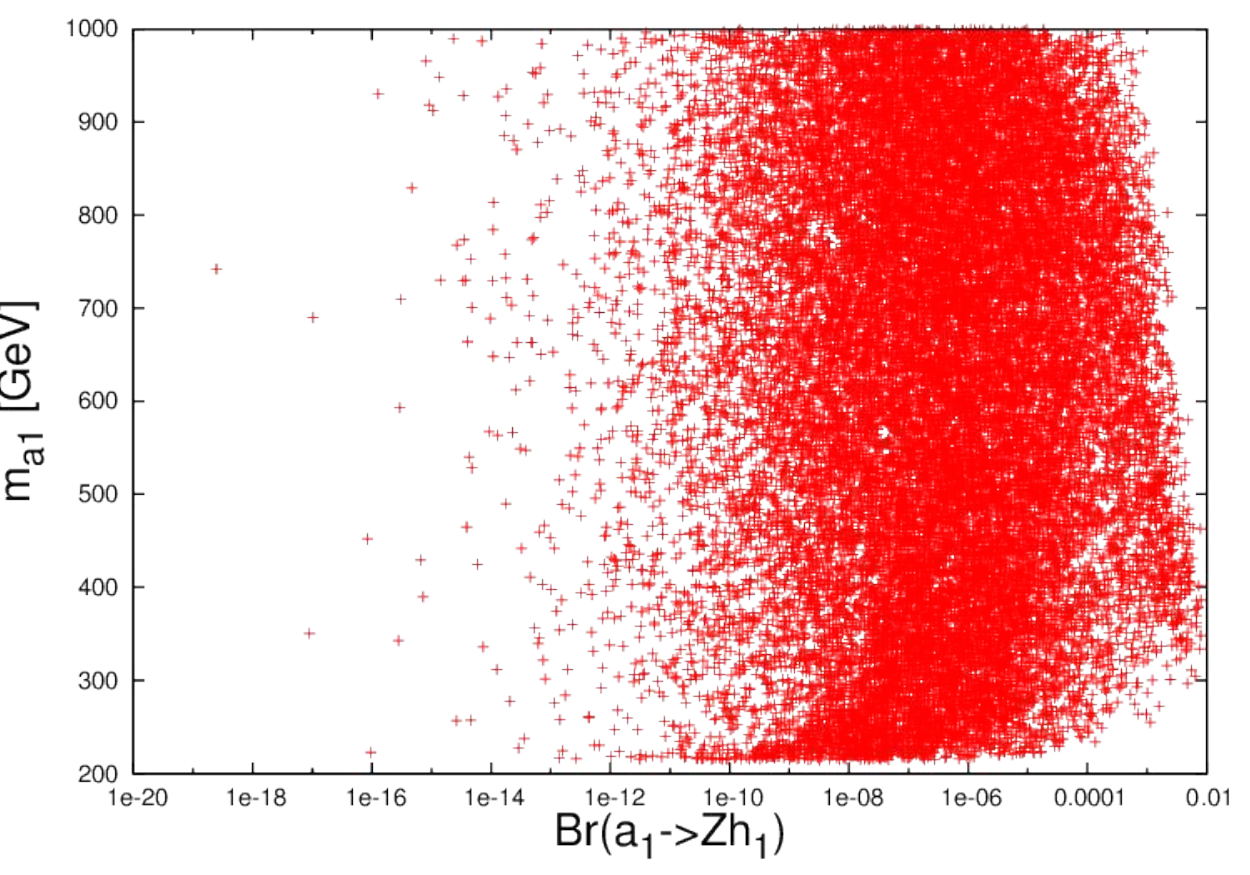}
 &\includegraphics[scale=0.50]{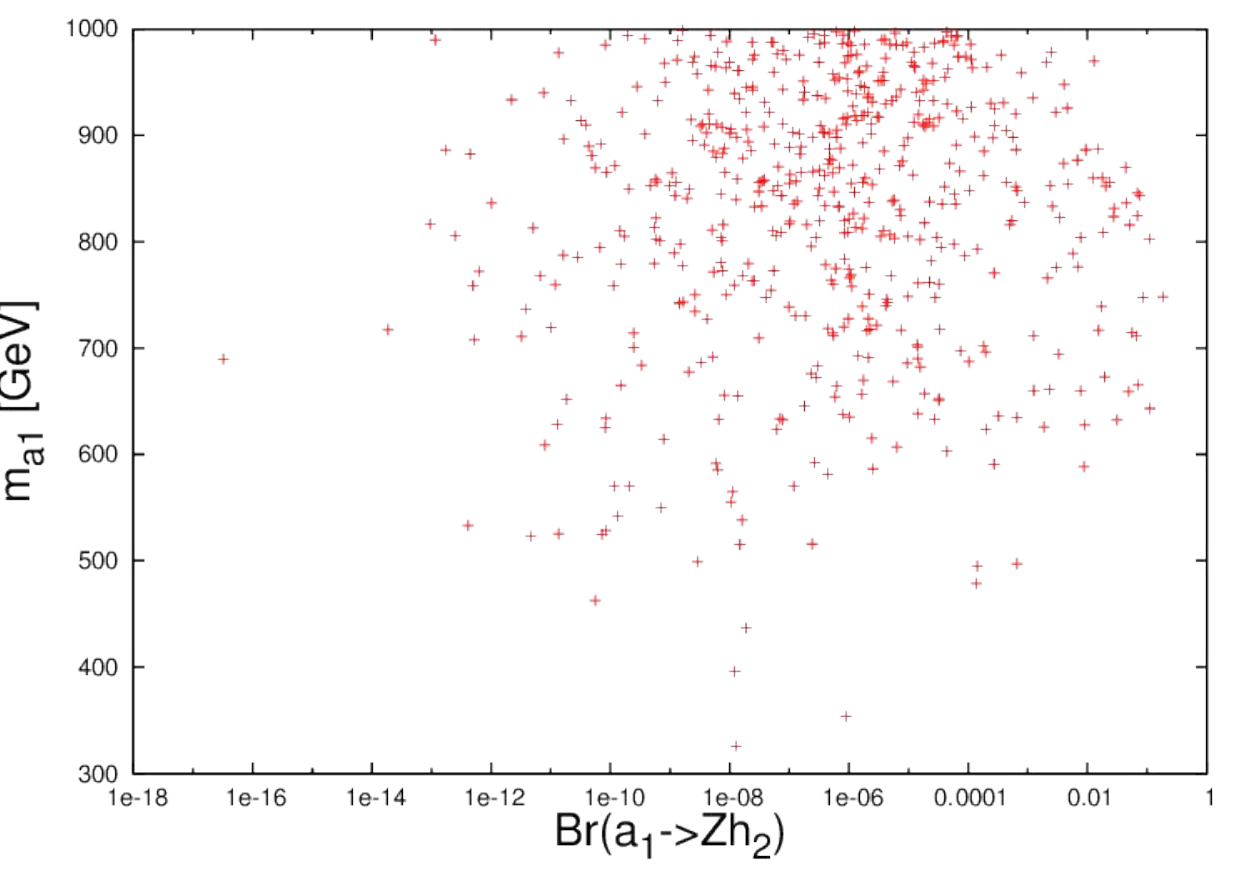}\\

 \end{tabular}
 \label{fig:mh1-ma1}
\caption{ The top panels show the relations between the lightest CP-even Higgs mass $m_{h_1}$ and each of
$m_{a_1}$, ${\rm Br}(h_1\to a_1a_1)$ and $m_{h_2}$. In the middle panels we plot
the second lightest CP-even Higgs mass $m_{h_2}$ against each of ${\rm Br}(h_2\to h_1h_1)$,  $m_{a_1}$ and ${\rm Br}(h_2\to a_1a_1)$.
Finally, the bottom panels show the relations between the lightest CP-odd Higgs mass $m_{a_1}$ and each of
${\rm Br}(a_1\to Zh_1)$ and ${\rm Br}(a_1\to Zh_2)$ .}
\end{figure}

\newpage
\section{Conclusions}
\label{sect:summa}
The NMSSM is phenomenologically richer than the MSSM in the Higgs and neutralino sectors due to the existence of the singlet 
Higgs superfield. We have explored large regions of the NMSSM parameter space through studying the decay modes of
the neutral lightest Higgses $h_1$, $h_2$ and $a_1$. We have noticed that large values of the $\tan\beta$ are still possible in
some points of the NMSSM parameter space although small and middle values are preferred. Furthermore, for all points selected in our
parameter space, which passed the experimental and the theoretical constraints, the $h_1$ is a SM-like Higgs due to 
the dominance of the Higgs doublet component of this lightest CP-even neutral physical Higgs. 

With regard to $h_2$ and $a_1$ decays, we have found two remarkable points. First, 
the branching fractions of $h_2\rightarrow b\bar b$ and of $a_1\rightarrow b\bar b$ can be dominant for all masses of $h_2$ and $a_1$
respectively due to the enhancement of Higgs couplings to down-type fermions. Second, the Higgs decays into the lightest neutralinos and charginos: $h_1\rightarrow\chi^0_1\chi^0_1$, $h_2\rightarrow\chi^0_1\chi^0_1$,
$h_2\rightarrow\chi^+_1\chi^-_1$, $a_1\rightarrow\chi^0_1\chi^0_1$, $a_1\rightarrow\chi^+_1\chi^-_1$ and also the decay modes $a_1\to \gamma\gamma$ and $a_1\to Z\gamma$
can be dominant in some area of the NMSSM parameter when $h_2$ and $a_1$ have predominant singlet component and a very weak doublet one.
In this case, the ideal collider to discover the Higgs 
singlet-like is $\gamma\gamma$ collider as the Higgs couplings to $\gamma\gamma$ at production level is enhanced.

Finally, the Higgs-to-Higgs decays play essential roles in the NMSSM. We have studied the decay channels that are kinematically allowed, which
are $h_{1, 2}\to a_1a_1$, $h_2\to h_1h_1$ and $a_1\to Zh_{1, 2}$. We have found that the decay modes $h_2\to a_1a_1$ and $h_2\to h_1h_1$ have
sizable branching ratios in large area of the NMSSM parameter space so they should be taken seriously when searching for the $h_2$ at the LHC.

\section*{Acknowledgments}
This work is funded by Taibah University, KSA.



\begin{thebibliography}{99}


\bibitem{ATLAS1-Higgs}
  G.~Aad {\it et al.} [ATLAS Collaboration],
  Phys.\ Lett.\ B {\bf 716}, 1 (2012).
\bibitem{ATLAS2-Higgs}
  G.~Aad {\it et al.} [ATLAS Collaboration],
  Phys.\ Lett.\ B {\bf 726}, 88 (2013).
\bibitem{CMS1-Higgs}
  S.~Chatrchyan {\it et al.} [CMS Collaboration],
  Phys.\ Lett.\ B {\bf 716}, 30 (2012).
\bibitem{CMS2-Higgs}
  S.~Chatrchyan {\it et al.} [CMS Collaboration],
  JHEP {\bf 1306}, 081 (2013).
\bibitem{SM1} S. Glashow, Nucl.\ Phys.\ {\bf 22}, 579 (1961).

\bibitem{SM2} S. Weinberg, Phys.\ Rev.\ Lett.\ {\bf 19}, 1264 (1967).

\bibitem{SM3}A. Salam, in “Elementary Particle Theory”, ed. N. Svartholm, Almqvist and Wiksells,
Stockholm (1969), p. 367.

\bibitem{SUSY1}P.~Ramond,  
  Phys.\ Rev.\  D {\bf 3}, 2415 (1971).
 \bibitem{SUSY2}A.~Neveu and J.~H.~Schwarz,
  Nucl.\ Phys.\  B {\bf 31}, 86 (1971).
 \bibitem{SUSY3}Yu.~A.~Golfand and E.~P.~Likhtman,
  JETP Lett.\  {\bf 13}, 323 (1971).
\bibitem{SUSY4}J.~L.~Gervais and B.~Sakita,
  Nucl.\ Phys.\  B {\bf 34}, 632 (1971).
\bibitem{SUSY5}D.~V.~Volkov and V.~P.~Akulov,
  Phys.\ Lett.\  B {\bf 46}, 109 (1973).
\bibitem{SUSY6}J.~Wess and B.~Zumino,
  Nucl.\ Phys.\  B {\bf 70}, 39 (1974).
\bibitem{SUSY7}A.~Salam and J.~A.~Strathdee,
  Nucl.\ Phys.\  B {\bf 76}, 477 (1974).
    
\bibitem{SUSY8}S.~P.~Martin, 
arXiv:hep-ph/9709356.
\bibitem{MSSMreview} A.~Djouadi,
  Phys.\ Rept.\  {\bf 459}, 1 (2008).
\bibitem{Kim:1983dt}
  J.~E.~Kim and H.~P.~Nilles,
  Phys.\ Lett.\  B {\bf 138}, 150 (1984).
\bibitem{Weinberg:1978ym}
  S.~Weinberg,
  Phys.\ Lett.\  B {\bf 82}, 387 (1979).
\bibitem{LlewellynSmith:1981yi}
  C.~H.~Llewellyn Smith and G.~G.~Ross,
  Phys.\ Lett.\  B {\bf 105}, 38 (1981).

\bibitem{NMSSMreviewed1}
M.~Maniatis, Int.\ J.\ Mod.\ Phys.\ A {\bf 25}, 3505 (2010).
\bibitem{NMSSMreviewed2}
U.~Ellwanger, C.~Hugonie and A.~M.~Teixeira, Phys.\ Rept.\ {\bf 496}, 1 (2010).

\bibitem{NMSSMrecent1}
  U.~Ellwanger,
  JHEP {\bf 1203}, 044 (2012).
  \bibitem{NMSSMrecent2}
  J.~F.~Gunion, Y.~Jiang and S.~Kraml,
  Phys.\ Lett.\ B {\bf 710}, 454 (2012).
  \bibitem{NMSSMrecent3}
  S.~F.~King, M.~Muhlleitner and R.~Nevzorov,
  Nucl.\ Phys.\ B {\bf 860}, 207 (2012).
  \bibitem{NMSSMrecent4}
  J.~J.~Cao, Z.~X.~Heng, J.~M.~Yang, Y.~M.~Zhang and J.~Y.~Zhu,
  JHEP {\bf 1203}, 086 (2012).
  \bibitem{NMSSMrecent5}
  D.~A.~Vasquez, G.~Belanger, C.~Boehm, J.~Da Silva, P.~Richardson and C.~Wymant,
  Phys.\ Rev.\ D {\bf 86}, 035023 (2012).
  \bibitem{NMSSMrecent6}
 U.~Ellwanger and C.~Hugonie,
  Adv.\ High Energy Phys.\  {\bf 2012}, 625389 (2012).
  \bibitem{NMSSMrecent7}
  R.~Benbrik, M.~Gomez Bock, S.~Heinemeyer, O.~Stal, G.~Weiglein and L.~Zeune,
  Eur.\ Phys.\ J.\ C {\bf 72}, 2171 (2012).
  \bibitem{NMSSMrecent8}
  J.~F.~Gunion, Y.~Jiang and S.~Kraml,
  Phys.\ Rev.\ D {\bf 86}, 071702 (2012).
  \bibitem{NMSSMrecent9}
  K.~J.~Bae, K.~Choi, E.~J.~Chun, S.~H.~Im, C.~B.~Park and C.~S.~Shin,
  JHEP {\bf 1211}, 118 (2012).
  \bibitem{NMSSMrecent10}
  K.~Agashe, Y.~Cui and R.~Franceschini,
  JHEP {\bf 1302}, 031 (2013).
  \bibitem{NMSSMrecent11}
  K.~Choi, S.~H.~Im, K.~S.~Jeong and M.~Yamaguchi,
  JHEP {\bf 1302}, 090 (2013).
  \bibitem{NMSSMrecent12}
  K.~Kowalska, S.~Munir, L.~Roszkowski, E.~M.~Sessolo, S.~Trojanowski and Y.~-L.~S.~Tsai,
  Phys.\ Rev.\ D {\bf 87}, 115010 (2013).
  \bibitem{NMSSMrecent13}
  S.~F.~King, M.~Mühlleitner, R.~Nevzorov and K.~Walz,
  Nucl.\ Phys.\ B {\bf 870}, 323 (2013).
  \bibitem{NMSSMrecent14}
  T.~Gherghetta, B.~von Harling, A.~D.~Medina and M.~A.~Schmidt,
  JHEP {\bf 1302}, 032 (2013).
   \bibitem{recentstudy1}
   M.~Badziak and C.~E.~M.~Wagner,
  JHEP {\bf 1702}, 050 (2017).  
 \bibitem{recentstudy2}
  J.~Beuria, U.~Chattopadhyay, A.~Datta and A.~Dey,
  JHEP {\bf 1704}, 024 (2017).
  \bibitem{recentstudy3}
  U.~Ellwanger,
  JHEP {\bf 1702}, 051 (2017).
  \bibitem{recentstudy4}
  S.~P.~Das and M.~Nowakowski,
  Phys.\ Rev.\ D {\bf 96}, no. 5, 055014 (2017).
  \bibitem{recentstudy5}
  P.~Drechsel, R.~Gröber, S.~Heinemeyer, M.~M.~Muhlleitner, H.~Rzehak and G.~Weiglein,
  Eur.\ Phys.\ J.\ C {\bf 77}, no. 6, 366 (2017).
  \bibitem{recentstudy6}
    J.~Cao, X.~Guo, Y.~He, P.~Wu and Y.~Zhang,
  Phys.\ Rev.\ D {\bf 95}, no. 11, 116001 (2017).  
\bibitem{NMHDECAY1} U. Ellwanger, J. F. Gunion and C. Hugonie, JHEP {\bf 0502}, 066 (2005).
\bibitem{NMHDECAY2} U. Ellwanger and C. Hugonie, Comput.\ Phys.\ Commun.\ {\bf 175}, 290 (2006).
\bibitem{NMSSMTools} See the Web site ``NMSSMTools: Tools for the
Calculation of the Higgs and Sparticle Spectrum in the
NMSSM: NMHDECAY, NMSPEC and NMGMSB", \\
http://www.th.u-psud.fr/NMHDECAY/nmssmtools.html.
\bibitem{NMSSM-Points} U.~Ellwanger, J.~F.~Gunion and C.~Hugonie, JHEP {\bf 0507}, 041 (2005).
\bibitem{Almarashi:2010jm}
  M.~M.~Almarashi and S.~Moretti,
  Eur.\ Phys.\ J.\  C {\bf 71}, 1618 (2011).
\bibitem{Almarashi:2011bf}
  M.~M.~Almarashi and S.~Moretti,
  Phys.\ Rev.\  D {\bf 84}, 015014 (2011).
\bibitem{Almarashi:2011hj}
  M.~M.~Almarashi and S.~Moretti,
  Phys.\ Rev.\  D {\bf 83}, 035023 (2011).
\bibitem{Almarashi:2011te}
   M.~M.~Almarashi and S.~Moretti,
  Phys.\ Rev.\  D {\bf 84}, 035009 (2011).
\bibitem{Almarashi:2011qq}
  M.~M.~Almarashi and S.~Moretti,
  Phys.\ Rev.\ D {\bf 85}, 017701 (2012).
  \bibitem{NMSSMreviewed3}
  M.~M.~Almarashi and S.~Moretti,
  arXiv:1205.1683 [hep-ph].
\bibitem{Lebedev} S. Andreas, O. Lebedev, S. R. Sanchez and A. Ringwald, JHEP {\bf 1008}, 003 (2010).


 
\end{thebibliography}
\end{document}